\documentclass[aps,showpacs,preprint]{revtex4}
\usepackage{graphicx}
\usepackage{epsfig}
\DeclareGraphicsExtensions{.eps}

\begin{document}
\title{Symmetries in Nuclei}
\author{P.~Van~Isacker}
\affiliation{
Grand Acc\'el\'erateur National d'Ions Lourds,
CEA/DSM--CNRS/IN2P3, B.P.~55027, F-14076 Caen Cedex 5, France}

\begin{abstract}
The use of dynamical symmetries or spectrum generating algebras
for the solution of the nuclear many-body problem is reviewed.
General notions of symmetry and dynamical symmetry
in quantum mechanics are introduced
and illustrated with simple examples
such as the SO(4) symmetry of the hydrogen atom
and the isospin symmetry in nuclei.
Two nuclear models,
the shell model and the interacting boson model,
are reviewed
with particular emphasis on their use of group-theoretical techniques.
\end{abstract}

\maketitle

\section{Introduction}
\label{s_intro}
In the {\it Oxford Dictionary of Current English}
symmetry is defined as the
`right correspondence of parts;
quality of harmony or balance
(in size, design etc.) between parts'.
The word is derived from Greek
where it has the meaning `with proportion' or `with order'.
In modern theories of physics
it has acquired a more precise meaning
but the general idea of seeking
to order physical phenomena still remains.
Confronted with the bewildering complexity
exhibited by the multitude of physical systems,
physicists attempt to extract
some simple regularities from observations,
and the fact that they can do so
is largely due to the presence of symmetries
in the laws of physics.
Although one can never hope
to explain all observational complexities
entirely on the basis of symmetry arguments alone,
these are nevertheless instrumental
in establishing correlations between
and (hidden) regularities in the data.

The mathematical theory of symmetry
is called group theory
and its origin dates back to the nineteenth century.
Of course, the notion of symmetry
is present implicitly in many mathematical studies
that predate the birth of group theory
and goes back even to the ancient Greeks, in particular Euclid.
It was, however, \'Evariste Galois
who perceived the importance of the group of permutations
to answer the question
whether the roots of a polynomial equation
can be algebraically represented or not.
(A readable summary of the solution of this problem
is given in the first chapter of Gilmore's book~\cite{Gilmore08}.)
In the process of solving
that long-standing mathematical problem
he invented group theory
as well as Galois theory
which studies the relation
between polynomials and groups.
The mathematical theory of groups
developed further throughout the nineteenth century
and made another leap forward in 1873
when Sophus Lie proposed
the concept of a Lie group and its associated Lie algebra.

For a long time it was assumed
that group theory was a branch of mathematics
without any application in the physical sciences.
This state of affairs changed
with the advent of quantum mechanics,
and it became clear that group theory
provides a powerful tool to understand
the structure of quantum systems from a unified perspective.
After the introduction of symmetry transformations in abstract spaces
(associated, for example, with isospin, flavor, color, etc.)
the role of group theory became even central.

The purpose of these lecture notes
is to introduce, explain and illustrate the concepts of symmetry and dynamical symmetry.
In Sect.~\ref{s_qm} a brief reminder is given
of the central role of symmetry in quantum mechanics
and of its relation with invariance and degeneracy.
There exist two standard examples
to illustrate the idea that symmetry implies degeneracy and {\it vice versa},
namely the hydrogen atom and the harmonic oscillator.
In Sect.~\ref{s_hydrogen} the first of them is analyzed in detail.
Section~\ref{s_dsym} describes the process of symmetry breaking
and, in particular, dynamical symmetry breaking
in the sense as it is used in these lecture notes.
This mechanism is illustrated in Sect.~\ref{s_isospin}
with a detailed example, namely isospin and its breaking in nuclei.
Sections~\ref{s_shell} and~\ref{s_ibm} then present
the nuclear shell model and the interacting boson model, respectively,
with a special emphasis on the symmetry techniques
that have been used in the context of these models.
Finally, in Sect.~\ref{s_conc}, a summary of these lecture notes is given. 

\section{Symmetry in quantum mechanics}
\label{s_qm}
The starting point of any discussion of symmetry
is that the laws of physics should be invariant
with respect to certain transformations of the reference frame,
such as a translation or rotation,
or a different choice of the origin of the time coordinate.
This observation leads to three fundamental conservation laws:
conservation of linear momentum, angular momentum and energy.
In some cases an additional space-inversion symmetry applies,
yielding another conserved quantity, namely parity.
In a relativistic framework
the above transformations on space and time
cannot be considered separately but become intertwined.
The laws of nature are then invariant
under the Lorentz transformations
which operate in four-dimensional space--time.

These transformations and their associated invariances
can be called `geometric'
in the sense that they are defined in space--time.
In quantum mechanics, an important extension of these concepts
is obtained by also considering transformations
that act in abstract spaces associated with intrinsic variables
such as spin, isospin (in atomic nuclei), flavor and color (of quarks) etc.
It is precisely these `intrinsic' invariances
which have lead to the preponderance of symmetry applications
in the quantum physics. 

To be more explicit,
consider a transformation acting on a physical system, that is,
an operation that transforms the coordinates $\bar r_i$
and the momenta $\bar p_i$
of the particles that constitute the system.
Such transformations are of a geometric nature.
For a discussion of symmetry in quantum-mechanical systems
this definition is too restrictive
and the appropriate generalization is to consider,
instead of the geometric transformations themselves,
the corresponding transformations in the Hilbert space
of quantum-mechanical states of the system.
The action of the geometric transformation on spin variables
({\it i.e.}, components of the spin vector)
is assumed to be identical to its action
on the components
of the angular momentum vector $\bar\ell=\bar r\wedge\bar p$.
Furthermore, it can be shown~\cite{Blaizot97}
that a correspondence exists
between the geometric transformations in physical space
and the transformations induced by it
in the Hilbert space of quantum-mechanical states.
This correspondence, however,
is not necessarily one-to-one;
that is only the case if the system is `bosonic'
(consists of any number of integer-spin bosons
and/or an even number of half-integer-spin fermions).
If the system is `fermionic'
(contains an odd number of fermions),
the correspondence is two-to-one
and the groups,
formed by the geometric transformations
and by the corresponding transformations
in the Hilbert space of quantum-mechanical states,
are not isomorphic but rather homomorphic.

No distinction is made in the following
between geometric and quantum-mechanical transformations;
all elements $g_i$ will be taken as operators
acting on the Hilbert space of quantum-mechanical states.

\subsection{Symmetry}
\label{ss_sym}
A time-independent Hamiltonian $H$
which commutes with the generators $g_k$
that form a Lie algebra G,
\begin{equation}
\forall g_k\in{\rm G}:
[H,g_k]=0,
\label{e_sym1}
\end{equation}
is said to have a symmetry G
or, alternatively, to be invariant under G.
The determination of operators $g_k$
that leave invariant the Hamiltonian
of a given physical system
is central to any quantum-mechanical description.
The reasons for this are profound
and can be understood from the correspondence
between geometrical and quantum-mechanical transformations.
It can be shown~\cite{Blaizot97}
that the transformations $g_k$
with the symmetry property~(\ref{e_sym1})
are induced by geometrical transformations
that leave unchanged
the corresponding classical Hamiltonian.
In this way the classical notion of a conserved quantity
is transcribed in quantum mechanics
in the form of the symmetry property~(\ref{e_sym1})
of the time-independent Hamiltonian.

\subsection{Degeneracy and state labeling}
\label{ss_deg}
A well-known consequence of a symmetry
is the occurrence of degeneracies
in the eigenspectrum of $H$.
Given an eigenstate $|\gamma\rangle$
of $H$ with energy $E$,
the condition~(\ref{e_sym1}) implies
that the states $g_k|\gamma\rangle$
all have the same energy,
\begin{equation}
H g_k|\gamma\rangle=
g_kH|\gamma\rangle=
Eg_k|\gamma\rangle.
\label{e_deg0}
\end{equation}
An arbitrary eigenstate of $H$
shall be written as $|\Gamma\gamma\rangle$,
where the first quantum number $\Gamma$
is different for states with different energies
and the second quantum number $\gamma$
is needed to label degenerate eigenstates.
The eigenvalues of a Hamiltonian
that satisfies~(\ref{e_sym1})
depend on $\Gamma$ only,
\begin{equation}
H|\Gamma\gamma\rangle=
E(\Gamma)|\Gamma\gamma\rangle,
\label{e_deg1}
\end{equation}
and, furthermore, the transformations $g_k$
do not admix states with different $\Gamma$,
\begin{equation}
g_k|\Gamma\gamma\rangle=
\sum_{\gamma'}a^\Gamma_{\gamma'\gamma}(k)
|\Gamma\gamma'\rangle.
\label{e_deg2}
\end{equation}

This simple discussion
of the consequences of a Hamiltonian symmetry
illustrates the relevance of group theory
in quantum mechanics.
Symmetry implies degeneracy
and eigenstates that are degenerate in energy
provide a Hilbert space
in which irreducible representations
of the symmetry group are constructed.
Consequently, the irreducible representations of a given group
directly determine the degeneracy structure
of a Hamiltonian with the symmetry associated to that group.

Eigenstates of $H$ can be denoted as $|\Gamma\gamma\rangle$
where the symbol $\Gamma$ labels
the irreducible representations of ${\rm G}$.
Note that the same irreducible representation
might occur more than once in the eigenspectrum of $H$
and, therefore, an additional multiplicity label $\eta$
should be introduced
to define a complete labeling of eigenstates
as $|\eta\Gamma\gamma\rangle$.
This label shall be omitted in the subsequent discussion.

A sufficient condition for a Hamiltonian
to have the symmetry property~(\ref{e_sym1})
is that it is a Casimir operator
which by definition commutes with all generators of the algebra.
The eigenequation~(\ref{e_deg1}) then becomes
\begin{equation}
C_m[{\rm G}]
|\Gamma\gamma\rangle=
E_m(\Gamma)
|\Gamma\gamma\rangle.
\label{e_deg3}
\end{equation}
In fact, all results remain valid
if the Hamiltonian is an analytic function
of Casimir operators of various orders.
The energy eigenvalues $E_m(\Gamma)$
are functions of the labels
that specify the irreducible representation $\Gamma$,
and are known for all classical Lie algebras~\cite{Wybourne74}.

These concepts can be illustrated with the example of the hydrogen atom
which is discussed in detail in the next section.

\section{The hydrogen atom}
\label{s_hydrogen}
The Hamiltonian for a particle of charge $-e$ and mass $m_{\rm e}$
in a Coulomb potential $e/r$ is given by
\begin{equation}
H_{\rm H}={\frac{p^2}{2m_{\rm e}}}-{\frac{e^2}{r}}=
{-\frac{\hbar^2}{2m_{\rm e}}}\nabla^2-{\frac{e^2}{r}}.
\label{e_hydham}
\end{equation}
This is taken here as a model Hamiltonian for the hydrogen atom.
The Hamiltonian is independent of the spin of the electron
which leads to a two-fold degeneracy of all states
corresponding to spin-up and spin-down.
Electron spin is ignored in the following
and the symmetry properties of the spatial part only
of the electron wave function are studied.

\begin{figure}
\centering
\includegraphics[width=11.5cm]{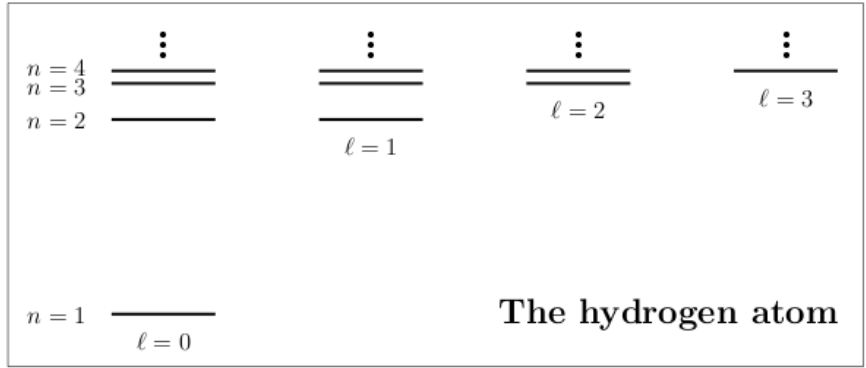}
\caption{The energy spectrum of the hydrogen atom.}
\label{f_hydspec}
\end{figure}

The solutions of the associated Schr\"odinger equation,
$H_{\rm H}\tilde\phi(\bar r)=E\tilde\phi(\bar r)$,
are well known from standard quantum mechanics.
The energies of the stationary states are
\begin{equation}
E(n)=-{\frac{m_{\rm e}e^4}{2\hbar^2n^2}}\equiv
-{\frac{R_{\rm H}}{n^2}},
\label{e_hyener}
\end{equation}
where $R_{\rm H}$ is the Rydberg constant
and $n$ the so-called principal quantum number.
The electron wave functions are
\begin{equation}
\tilde\phi_{n\ell m_\ell}(r,\theta,\varphi)=
\tilde R_{n\ell}(r)
Y_{\ell m_\ell}(\theta,\varphi),
\label{e_hywave}
\end{equation}
with $\tilde R_{n\ell}(r)$ and $Y_{\ell m_\ell}(\theta,\varphi)$
known functions\footnote{The notation with a tilde is used
to distinguish the radial part of the wave function
from the $R_{n\ell}(r)$ for the harmonic oscillator
that will be encountered in Sect.~\ref{s_shell}.}.
The $Y_{\ell m_\ell}(\theta,\varphi)$ are spherical harmonics
which occur for any central potential with spherical symmetry.
The $\tilde R_{n\ell}(r)$ are radial wave functions
whose exact form is not of concern here.
The solution of the differential equation
$H_{\rm H}\tilde\phi(\bar r)=E\tilde\phi(\bar r)$
also leads to the conditions
\begin{equation}
n=1,2,\dots,
\qquad
\ell=0,1,\dots,n-1,
\qquad
m_\ell=-\ell,-\ell+1,\dots,+\ell.
\end{equation}
The energy spectrum of the hydrogen atom
is shown in Fig.~\ref{f_hydspec}.
The energy eigenvalues $E(n)$ only depend on $n$
and not on $\ell$ or $m_\ell$.
A given level with energy $E(n)$ is thus $n^2$-fold degenerate since
\begin{equation}
\sum_{\ell=0}^{n-1}(2\ell+1)=n^2.
\end{equation}
The nature of this degeneracy
will be explained using symmetry arguments
and, in addition, it will be shown that the entire spectrum
can be determined with algebraic methods
without recourse to boundary conditions of differential equations.

The Hamiltonian of the hydrogen atom
is rotationally [or SO(3)] invariant.
This is obvious on intuitive grounds
since the properties of the hydrogen atom do not change under rotation.
Formally, it follows from the following commutation property:
\begin{equation}
[H_{\rm H},L_\mu]=0,
\end{equation}
where $L_\mu$ are the components
of the angular momentum operator\footnote{Throughout these lecture notes,
small letters are normally reserved for operators
associated with a single particle
and capital letters for operators summed over many particles.
This section deals with one-particle operators
but for clarity's sake
it is important to distinguish between operators,
which shall be denoted in this section by capital letters $L$, $P$,\dots,
and their associated labels,
which shall be denoted by corresponding small letters $\ell$, $p$,\dots.},
$\bar L=(\bar r\wedge\bar p)=-i\hbar(\bar r\wedge\bar\nabla)$.
It is of interest to look more closely
at the origin of the vanishing commutator
between $H_{\rm H}$ and $L_\mu$.
The Hamiltonian of the hydrogen atom
consists of two parts, kinetic and potential,
and {\em both} commute with $L_\mu$ since
\begin{equation}
[\nabla^2,L_\mu]=0,
\qquad
[r^{-1},L_\mu]=0,
\end{equation}
where use is made of commutation relations like
\begin{equation}
[\bar\nabla,r^k]=k\,r^{k-2}\bar r,
\quad
[\nabla^2,\bar r]=2\bar\nabla.
\end{equation}
Since the components $L_\mu$ form an SO(3) algebra,
\begin{equation}
[L_\mu,L_\nu]=i\hbar\sum_{\rho=1}^3\epsilon_{\mu\nu\rho}L_\rho,
\end{equation}
and since $L_\mu$ commutes with $H_{\rm H}$,
one concludes that the Hamiltonian of the hydrogen atom
has an SO(3) symmetry.
This explains part of the observed degeneracy,
namely, levels with a given $\ell$ are $(2\ell+1)$-fold degenerate.

\begin{figure}
\centering
\includegraphics[width=6.5cm]{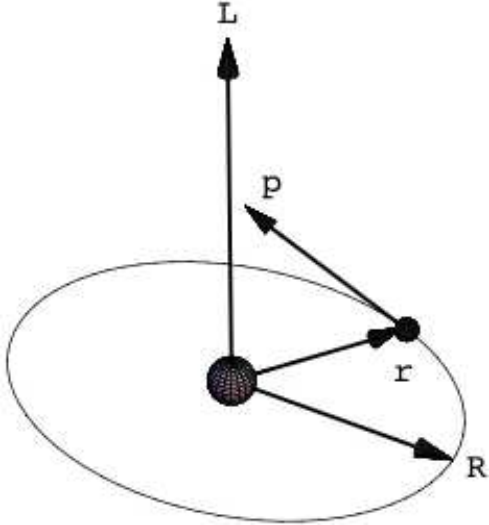}
\caption{The angular momentum vector $\bar L$
and the Runge--Lenz vector $\bar R$
in the classical Kepler problem
of a planet orbiting the sun.}
\label{f_kepler}
\end{figure}

To understand the origin of the {\em complete} degeneracy
of the hydrogen spectrum,
it is instructive to consider first
the Kepler problem of the motion of a single planet around the sun
which is the classical analogue of the hydrogen atom.
Besides angular momentum,
there is another conserved quantity
because there is no precession of the planetary orbit,
that is, the major axis of its elliptic trajectory is fixed.
In contrast to the conservation of angular momentum
which is valid for {\em all} central potentials,
the absence of precession is a specific property
of the Newtonian $1/r$ potential.
The associated conserved quantity
is known from classical mechanics,
\begin{equation}
\bar R_{\rm cl}=
{\frac{\bar p\wedge\bar L}{m_{\rm e}}}-
e^2{\frac{\bar r}{r}}.
\end{equation}
This vector is known as the Runge--Lenz (or also Lenz--Pauli) vector
and its three components are conserved for a $1/r$ potential,
that is, not only its direction (along the major axis of the orbit)
but also its magnitude is conserved (see Fig.~\ref{f_kepler}).
The latter property follows from the relation
\begin{equation}
\bar R_{\rm cl}^2=e^4+{\frac{2E}{m_{\rm e}}}\bar L^2,
\label{e_rule}
\end{equation}
which shows that $\bar R_{\rm cl}^2$ can be expressed
in terms of the energy and the angular momentum,
both of which are conserved.

The construction of the quantum-mechanical equivalent
of the Runge--Lenz vector
is done in the usual way and yields
\begin{equation}
\bar R'=
-{\frac{\hbar^2}{2m_{\rm e}}}
[\bar\nabla\wedge(\bar r\wedge\bar\nabla)-
(\bar r\wedge\bar\nabla)\wedge\bar\nabla]-
e^2{\frac{\bar r}{r}}.
\end{equation}
The relation~(\ref{e_rule}) between the energy
and the moduli of the angular momentum and Runge--Lenz vectors
converts to
\begin{equation}
\bar R'^2=e^4+{\frac{2H_{\rm H}}{m_{\rm e}}}\left(\bar L^2+\hbar^2\right).
\label{e_hydrel1}
\end{equation}
From the classical analysis
one expects $R'_\mu$ to commute with $H_{\rm H}$,
\begin{equation}
[H_{\rm H},R'_\mu]=0,
\end{equation}
which is indeed confirmed through explicit calculation.
Unlike in the case of the angular momentum, however,
it is only the entire Hamiltonian
which commutes with the Runge--Lenz vector,
and {\em not} the kinetic and potential parts separately since
\begin{equation}
{-\frac{\hbar^2}{2m_{\rm e}}}[\nabla^2,R'_\mu]=
e^2[r^{-1},R'_\mu]=
{\frac{\hbar^2e^2}{m_{\rm e}}}
\left[
{\frac{1}{r}}\nabla_\mu-
{\frac{r_\mu}{r^3}}(1+\bar r\cdot\bar\nabla)\right]\neq0.
\end{equation}
Just as in the classical Kepler problem
with its exceptional precessionless orbits,
one finds that the commutator with the Runge--Lenz vector
vanishes for a $1/r$ potential but not in general.

It is now established
that both vectors $\bar L$ and $\bar R'$
commute with the Hamiltonian of the hydrogen atom
and hence are constants of motion,
but the symmetry of the system still needs to be determined.
This can be done from the commutation relations
among $L_\mu$ and $R'_\mu$ which read
\begin{equation}
[L_\mu,R'_\nu]=i\hbar\sum_{\rho=1}^3\epsilon_{\mu\nu\rho}R'_\rho,
\qquad
[R'_\mu,R'_\nu]=i\hbar\sum_{\rho=1}^3\epsilon_{\mu\nu\rho}
{\frac{-2H_{\rm H}}{m_{\rm e}}}L_\rho,
\end{equation}
together with the SO(3) relations among $L_\mu$.
Since the commutation relations among $R'_\mu$
do {\em not} give back $L_\rho$,
one cannot claim that $L_\mu$ and $R'_\mu$
form a Lie algebra.
In the space of eigenvectors
corresponding to a single, negative eigenvalue,
the following alternative operators can be introduced:
\begin{equation}
R_\mu=\sqrt{\frac{m_{\rm e}}{-2H_{\rm H}}}R'_\mu.
\end{equation}
In general, the square-root of an operator
has problematic properties
but not in this case
since it acts in a space of constant eigenvalue.
Note also that one may rely here on the fact
that neither $L_\mu$ nor $R_\mu$ or $R'_\mu$
can connect to states with a different energy eigenvalue,
since they all commute with $H_{\rm H}$.
The commutation relations
among $L_\mu$ and $R_\mu$ now close,
\begin{equation}
[L_\mu,R_\nu]=i\hbar\sum_{\rho=1}^3\epsilon_{\mu\nu\rho}R_\rho,
\qquad
[R_\mu,R_\nu]=i\hbar\sum_{\rho=1}^3\epsilon_{\mu\nu\rho}L_\rho,
\end{equation}
and the algebra consisting of $L_\mu$ and $R_\mu$
can be identified with SO(4),
associated with the group of rotations in four dimensions.
The relation~(\ref{e_hydrel1}) between the Hamiltonian
and the conserved quantities $\bar L^2$ and $\bar R^2$
can be rewritten as
\begin{equation}
H_{\rm H}=
-{\frac{\hbar^2R_{\rm H}}{\bar L^2+\bar R^2+\hbar^2}}.
\label{e_hydrel2}
\end{equation}
The operator occurring at the right-hand side of this identity,
$\bar L^2+\bar R^2$,
can be identified with $C_2[{\rm SO}(4)]$,
the quadratic Casimir operator of SO(4).
The hydrogen atom provides thus a simple example
in which a Hamiltonian can be written
in terms of the Casimir operator of its symmetry algebra.

In general, if the symmetry group of a Hamiltonian is determined,
its degeneracy structure follows automatically
from the irreducible representations
which can be looked up in monographs on group theory.
In the case of SO(4) the analysis can be worked out
with simple methods
by converting to the operators
\begin{equation}
P_\mu={\frac 1 2}\left(L_\mu+R_\mu\right),
\qquad
Q_\mu={\frac 1 2}\left(L_\mu-R_\mu\right),
\end{equation}
in terms of which the commutation relations become
\begin{equation}
[P_\mu,P_\nu]=i\hbar\sum_{\rho=1}^3\epsilon_{\mu\nu\rho}P_\rho,
\quad
[Q_\mu,Q_\nu]=i\hbar\sum_{\rho=1}^3\epsilon_{\mu\nu\rho}Q_\rho,
\quad
[P_\mu,Q_\nu]=0.
\end{equation}
The components $P_\mu$ commute with $Q_\nu$
and, furthermore, each set separately forms an SO(3) algebra.
This, in fact, proves the isomorphism
${\rm SO}(4)\simeq{\rm SO}(3)\otimes{\rm SO}(3)$.
Instead of relying on SO(4) representation theory,
one can therefore use well-known results from SO(3).
Since the operators $P^2$, $P_z$, $Q^2$ and $Q_z$
commute with each other,
and since they all commute with $H_{\rm H}$,
they form a (complete) set of commuting operators. 
The eigenstates of $H_{\rm H}$
can then be labeled with the eigenvalues
of the operators in this set
and, in particular, with $p(p+1)\hbar^2$ and $q(q+1)\hbar^2$,
the eigenvalues of the operators $\bar P^2$ and $\bar Q^2$.
The allowed values of the labels $p$ and $q$
are those of angular momentum, integer or half-integer,
and for each value of $p$ ($q$)
there are $2p+1$ ($2q+1$) allowed substates. 
Furthermore, eigenstates of $H_{\rm H}$ necessarily have $p=q$
because the angular momentum and the Runge--Lenz vectors
are orthogonal, $\bar L\cdot\bar R=0$,
which implies
\begin{equation}
\left(\bar L+\bar R\right)^2=
\left(\bar L-\bar R\right)^2
\Rightarrow
\bar P^2=\bar Q^2
\Rightarrow
p(p+1)=q(q+1).
\end{equation}
The allowed energy eigenvalues
are now immediately obtained from~(\ref{e_hydrel2})
since the operator $\bar L^2+\bar R^2=2\bar P^2+2\bar Q^2$
has the eigenvalue $4p(p+1)\hbar^2$,
\begin{equation}
E(p)=-{\frac{\hbar^2R_{\rm H}}{4p(p+1)\hbar^2+\hbar^2}}=
-{\frac{R_{\rm H}}{(2p+1)^2}},
\qquad
p=0,{\textstyle{\frac 1 2}},1,\dots.
\end{equation}
This coincides with the result~(\ref{e_hyener})
obtained from the standard quantum-mechanical derivation.

The hydrogen atom provides
a beautiful application of symmetry.
The degeneracies observed in the energy spectrum
are higher than what is obtained from just rotational invariance.
This requires the existence of a larger symmetry
which is indeed found to be the case.
Another illustration of this principle
is provided by the spectrum of the harmonic oscillator
in which case the underlying symmetry turns out to be U(3)~\cite{Moshinsky69}.

A final comment concerns the method
followed here to determine the eigenspectrum
of the hydrogen atom.
The standard way to do so
is to solve the time-independent Schr\"odinger equation
and to find the allowed values of the various quantum numbers
from boundary conditions on the eigenfunctions.
The procedure followed here is entirely different
and exclusively based on the knowledge
of a set of constants of motion which commute with the Hamiltonian,
together with their mutual commutation relations.
A crucial feature is that the Hamiltonian
can be expressed in terms of the Casimir operator of the symmetry algebra.
Although elegant and compact,
the method itself does not provide
an expression for the wave functions of stationary states.
This `algebraic' solution method
of the problem of the hydrogen atom
was proposed by Pauli in 1926~\cite{Pauli26}.

\section{Dynamical symmetry breaking}
\label{s_dsym}
The concept of a dynamical symmetry
for which (at least) two algebras ${\rm G}_1$ and ${\rm G}_2$
with ${\rm G}_1\supset{\rm G}_2$ are needed
can now be introduced.
The eigenstates of a Hamiltonian $H$
with symmetry ${\rm G}_1$
are labeled as $|\Gamma_1\gamma_1\rangle$.
But, since ${\rm G}_1\supset{\rm G}_2$,
a Hamiltonian with ${\rm G}_1$ symmetry
necessarily must also have a symmetry ${\rm G}_2$
and, consequently, its eigenstates can also be labeled as 
$|\Gamma_2\gamma_2\rangle$.
Combination of the two properties leads to the eigenequation
\begin{equation}
H|\Gamma_1\eta_{12}\Gamma_2\gamma_2\rangle=
E(\Gamma_1)|\Gamma_1\eta_{12}\Gamma_2\gamma_2\rangle,
\label{e_dsym1}
\end{equation}
where the role of $\gamma_1$
is played by $\eta_{12}\Gamma_2\gamma_2$.
The irreducible representation $\Gamma_2$
may occur more than once in $\Gamma_1$,
and hence an additional quantum number $\eta_{12}$
is needed to uniquely label the states.
Because of ${\rm G}_1$ symmetry,
eigenvalues of $H$ depend on $\Gamma_1$ only.

In many examples in physics (several are discussed below),
the condition of ${\rm G}_1$ symmetry is too strong
and a {\em possible} breaking of the ${\rm G}_1$ symmetry
can be imposed via the Hamiltonian
\begin{equation}
H'=
\kappa_1C_{m_1}[{\rm G}_1]+
\kappa_2C_{m_2}[{\rm G}_2],
\end{equation}
which consists of a combination of Casimir operators
of ${\rm G}_1$ {\em and} ${\rm G}_2$.
The symmetry properties of the Hamiltonian $H'$
are now as follows.
Since $[H',g_k]=0$
for all $g_k$ in ${\rm G}_2$,
$H'$ is invariant under ${\rm G}_2$.
The Hamiltonian $H'$,
since it contains  $C_{m_2}[{\rm G}_2]$,
does not commute, in general, with all elements of ${\rm G}_1$
and for this reason the ${\rm G}_1$ symmetry is broken.
Nevertheless,
because $H'$ is a combination of Casimir operators
of ${\rm G}_1$ and ${\rm G}_2$,
its eigenvalues can be obtained in closed form,
\begin{equation}
H'
|\Gamma_1\eta_{12}\Gamma_2\gamma_2\rangle=
\left[
\kappa_1E_{m_1}(\Gamma_1)+
\kappa_2E_{m_2}(\Gamma_2)\right]
|\Gamma_1\eta_{12}\Gamma_2\gamma_2\rangle.
\end{equation}
The conclusion is thus that,
although $H'$ is not invariant under ${\rm G}_1$,
its eigenstates are the same
as those of $H$ in~(\ref{e_dsym1}).
The Hamiltonian $H'$
is said to have ${\rm G}_1$ as a dynamical symmetry.
The essential feature is that,
although the eigenvalues of $H'$
depend on $\Gamma_1$ {\em and} $\Gamma_2$
(and hence ${\rm G}_1$ is not a symmetry),
the eigenstates do not change
during the breaking of the ${\rm G}_1$ symmetry.
As the generators of ${\rm G}_2$
are a subset of those of ${\rm G}_1$,
the dynamical symmetry breaking splits
but does not admix the eigenstates.
A convenient way of summarizing
the symmetry character of $H'$
and the ensuing classification of its eigenstates
is as follows:
\begin{equation}
\begin{array}{ccc}
{\rm G}_1&\supset&{\rm G}_2\\
\downarrow&&\downarrow\\
\Gamma_1&&\eta_{12}\Gamma_2
\end{array}.
\end{equation}
This equation indicates the larger algebra ${\rm G}_1$
(sometimes referred to as the dynamical algebra
or spectrum generating algebra)
and the symmetry algebra ${\rm G}_2$,
together with their associated labels with possible multiplicities.

Many concrete examples exist in physics
of the abstract idea of dynamical symmetry.
Perhaps the best known in nuclear physics
concerns isospin symmetry
and its breaking by the Coulomb interaction
which is discussed in the next section.

\section{Isospin symmetry}
\label{s_isospin}
The starting point in the discussion of isospin symmetry
is the observation
that the masses of the neutron and proton are very similar,
$m_{\rm n}c^2=939.55$~MeV and $m_{\rm p}c^2=938.26$~MeV,
and that both have a spin of ${\frac 1 2}$.
Furthermore, experiment shows
that, if one neglects the contribution
of the electromagnetic interaction,
the forces between two neutrons
are about the same 
as those between two protons.
More precisely, the strong nuclear force between two nucleons
with anti-parallel spins
is found to be (approximately) independent
of whether they are neutrons or protons.
This indicates the existence
of a symmetry of the strong interaction,
and isospin is the appropriate formalism
to explore the consequences of that symmetry in nuclei.
The equality of the masses and the spins of the nucleons
is not sufficient for isospin symmetry to be valid
and the charge independence of the nuclear force
is equally important.
This point was emphasized by Wigner~\cite{Wigner37}
who defined isospin for complex nuclei as we know it today
and who also coined the name of `isotopic spin'.

Because of the near-equality
of the masses and of the interactions between nucleons,
the Hamiltonian of the nucleus is (approximately) invariant
with respect to transformations
between neutron and proton states.
For one nucleon, these can be defined
by introducing the abstract space spanned by the two vectors
\begin{equation}
|{\rm n}\rangle=
\left[\begin{array}{c}
1\\0
\end{array}\right],
\qquad
|{\rm p}\rangle=
\left[\begin{array}{c}
0\\1
\end{array}\right].
\end{equation}
The most general transformation among these states
(which conserves their normalization)
is a unitary $2\times2$ matrix.
A matrix close to the identity can be represented as
\begin{equation}
\left[\begin{array}{ccc}
1+\epsilon_{11}&&\epsilon_{12}\\
\epsilon_{21}&&1+\epsilon_{22}
\end{array}\right],
\end{equation}
where the $\epsilon_{ij}$ are infinitesimal complex numbers.
Unitarity imposes the relations
\begin{equation}
\epsilon_{11}+\epsilon_{11}^*=
\epsilon_{22}+\epsilon_{22}^*=
\epsilon_{12}+\epsilon_{21}^*=0.
\end{equation}
An additional condition is found
by requiring the determinant of the unitary matrix
to be equal to $+1$,
\begin{equation}
\epsilon_{11}+\epsilon_{22}=0,
\end{equation}
which removes the freedom to make
a simultaneous and identical change of phase
for the neutron and the proton.
The infinitesimal, physical transformations
between a neutron and a proton
can therefore be parametrized as
\begin{equation}
\left[\begin{array}{ccc}
1-{\frac 1 2}i\epsilon_z&&-{\frac 1 2}i(\epsilon_x-i\epsilon_y)\\
-{\frac 1 2}i(\epsilon_x+i\epsilon_y)&&1+{\frac 1 2}i\epsilon_z
\end{array}\right],
\end{equation}
which includes a conventional factor $-i/2$
and where the $\{\epsilon_x,\epsilon_y,\epsilon_z\}$
now are infinitesimal real numbers.
This can be rewritten
in terms of the Pauli spin matrices as
\begin{equation}
\left[\begin{array}{rcr}
1&&0\\0&&1
\end{array}\right]
-{\frac 1 2}i\epsilon_x
\left[\begin{array}{rcr}
0&&1\\1&&0
\end{array}\right]
-{\frac 1 2}i\epsilon_y
\left[\begin{array}{rcr}
0&&-i\\i&&0
\end{array}\right]
-{\frac 1 2}i\epsilon_z
\left[\begin{array}{rcr}
1&&0\\0&&-1
\end{array}\right].
\end{equation}
The infinitesimal transformations between a neutron and a proton
can thus be written in terms of the three operators
\begin{equation}
t_x\equiv
{\frac 1 2}
\left[\begin{array}{rcr}
0&&1\\1&&0
\end{array}\right],
\qquad
t_y\equiv
{\frac 1 2}
\left[\begin{array}{rcr}
0&&-i\\i&&0
\end{array}\right],
\qquad
t_z\equiv
{\frac 1 2}
\left[\begin{array}{rcr}
1&&0\\0&&-1
\end{array}\right],
\end{equation}
which satisfy {\em exactly} the same commutation relations
as the angular momentum operators.
The action of the $t_\mu$ operators on a nucleon state
is easily found from its matrix representation.
For example,
\begin{equation}
t_z|{\rm n}\rangle\equiv
{\frac 1 2}
\left[\begin{array}{rcr}
1&&0\\0&&-1
\end{array}\right]
\left[\begin{array}{c}
1\\0
\end{array}\right]=
{\frac 1 2}
|{\rm n}\rangle,
\quad
t_z|{\rm p}\rangle\equiv
{\frac 1 2}
\left[\begin{array}{rcr}
1&&0\\0&&-1
\end{array}\right]
\left[\begin{array}{c}
0\\1
\end{array}\right]=
-{\frac 1 2}
|{\rm p}\rangle,
\end{equation}
which shows that $e(1-2t_z)/2$ is the charge operator. 
Also, the combinations
$t_\pm\equiv t_x\pm it_y$ can be introduced,
which satisfy the commutation relations
\begin{equation}
[t_z,t_\pm]=\pm t_\pm,
\qquad
[t_+,t_-]=2t_z,
\end{equation}
and play the role of raising and lowering operators
since
\begin{equation}
t_-|{\rm n}\rangle=|{\rm p}\rangle,
\qquad
t_+|{\rm n}\rangle=0,
\qquad
t_-|{\rm p}\rangle=0,
\qquad
t_+|{\rm p}\rangle=|{\rm n}\rangle.
\end{equation}

This proves the formal equivalence between spin and isospin,
and all results familiar from angular momentum
can now be readily transposed to the isospin algebra.
For a many-nucleon system (such as a nucleus)
a total isospin $T$ and its $z$ projection $M_T$ can be defined
which results from the coupling of the individual isospins,
just as this can be done for the nucleon spins.  
The appropriate isospin operators are
\begin{equation}
T_\mu=\sum_{k=1}^At_\mu(k),
\end{equation}
where the sum is over all the nucleons in the nucleus.

If, in first approximation,
the Coulomb interaction between the protons is neglected
and, furthermore, if it is assumed
that the strong interaction does not distinguish
between neutrons and protons,
the resulting nuclear Hamiltonian $H$
is isospin invariant.
Explicitly, invariance under the isospin algebra
${\rm SU}(2)\equiv\{T_z,T_\pm\}$
follows from
\begin{equation}
[H,T_z]=[H,T_\pm]=0.
\end{equation}
As a consequence of these commutation relations,
the many-particle eigenstates of $H$
have good isospin symmetry.
They can be classified as $|\eta TM_T\rangle$
where $T$ is the total isospin of the nucleus
obtained from the coupling
of the individual isospins ${\frac 1 2}$ of all nucleons,
$M_T$ is its projection on the $z$ axis in isospin space,
$M_T=(N-Z)/2$
and $\eta$ denotes all additional quantum numbers.
If isospin were a true symmetry,
all states $|\eta TM_T\rangle$ with $M_T=-T,-T+1,\dots,+T$,
and with the same $T$
(and identical other quantum numbers $\eta$),
would be degenerate in energy;
for example, neutron and proton
would have exactly the same mass.
States with the same $\eta T$ but different $M_T$
(and hence in different nuclei)
are referred to as isobaric analogue states.

\subsection{The isobaric multiplet mass equation}
\label{ss_imme}
The Coulomb interaction between the protons
destroys the equivalence between the nucleons
and hence breaks isospin symmetry.
The main effect of the Coulomb interaction
is a {\em dynamical} breaking of isospin symmetry.
This can be shown by rewriting the Coulomb interaction,
\begin{equation}
V_{\rm C}=
\sum_{k<l}^A
\left({\frac 1 2}-t_z(k)\right)
\left({\frac 1 2}-t_z(l)\right)
{\frac{e^2}{|\bar r_k-\bar r_l|}},
\end{equation}
as a sum of isoscalar, isovector and isotensor parts
\begin{equation}
V_{\rm C}=
\sum_{k<l}^A\sum_{t=0,1,2}
V^{(t)}_0(k,l),
\end{equation}
with
\begin{eqnarray}
V^{(0)}_0(k,l)&=&
\left({\frac 1 4}-\sqrt{\frac 1 3}
\left[\bar t(k)\times\bar t(l)\right]^{(0)}_0\right)
{\frac{e^2}{|\bar r_k-\bar r_l|}},
\nonumber\\
V^{(1)}_0(k,l)&=&
-{\frac 1 2}\left[t_z(k)+t_z(l)\right]
{\frac{e^2}{|\bar r_k-\bar r_l|}},
\nonumber\\
V^{(2)}_0(k,l)&=&
\sqrt{\frac 2 3}\left[\bar t(k)\times\bar t(l)\right]^{(2)}_0
{\frac{e^2}{|\bar r_k-\bar r_l|}},
\end{eqnarray}
where the coupling is carried out in isospin.
The Wigner--Eckart theorem in isospin space
allows to factor out the $M_T$ dependence
of any diagonal matrix element according to
\begin{equation}
\langle\eta TM_T|
\sum_{k<l}^AV^{(t)}_0(k,l)
|\eta TM_T\rangle=
\langle TM_T\;t0|TM_T\rangle
\langle\eta T\|
\sum_{k<l}^AV^{(t)}(k,l)
\|\eta T\rangle,
\end{equation}
where $\langle TM_T\;t0|TM_T\rangle$ is a Clebsch--Gordan coefficient
associated with ${\rm SU}(2)\supset{\rm SO}(2)$.
From the explicit expressions for these coefficients,
\begin{eqnarray}
\langle TM_T\;00|TM_T\rangle&=&1,
\qquad
\langle TM_T\;10|TM_T\rangle=
{\frac{M_T}{\sqrt{T(T+1)}}},
\nonumber\\
\langle TM_T\;20|TM_T\rangle&=&
{\frac{3M_T^2-T(T+1)}{\sqrt{T(T+1)(2T-1)(2T+3)}}},
\end{eqnarray}
one concludes that the $M_T$ dependence
of the diagonal matrix elements of the Coulomb interaction
is at most quadratic.
If the off-diagonal, isospin mixing matrix elements
of $V_{\rm C}$ are neglected,
it can then be represented as
\begin{equation}
V_{\rm C}\approx\kappa_0+\kappa_1T_z+\kappa_2T_z^2,
\label{e_vcapp}
\end{equation}
for some particular coefficients
$\kappa_0$, $\kappa_1$ and $\kappa_2$
which, according to the preceding discussion,
depend on the isospin $T$ and other quantum numbers $\eta$.
This can be viewed as a dynamical symmetry breaking
of the type
\begin{equation}
\begin{array}{ccc}
{\rm SU}(2)&\supset&{\rm SO}(2)\equiv\{T_z\}\\
\downarrow&&\downarrow\\
T&&M_T
\end{array}.
\end{equation}
The Hamiltonian~(\ref{e_vcapp}) splits but does not admix
the eigenstates $|\eta TM_T\rangle$ with $M_T=-T,-T+1,\dots,+T$,
and has the eigenspectrum
\begin{equation}
E(M_T)=\kappa_0+\kappa_1M_T+\kappa_2M_T^2.
\end{equation}

The expansion in $T_z$ is but an approximation
to the true Coulomb interaction;
it represents the diagonal part of it,
with the $T$-mixing isovector and isotensor parts
being neglected.
In that approximation
isospin remains a good quantum number.
The excitation spectra of the different nuclei
belonging to the same isospin multiplet
(with the same $T$ but different $M_T$)
are identical
but their ground states
do not have the same binding energy.
The energy formula in $M_T$
was derived by Wigner~\cite{Wigner57}
who introduced the name of
isobaric multiplet mass equation (IMME).
Many experimental examples
of nuclear isospin multiplets are known at present~\cite{Britz98}.

\begin{figure}
\centering
\includegraphics[width=10cm]{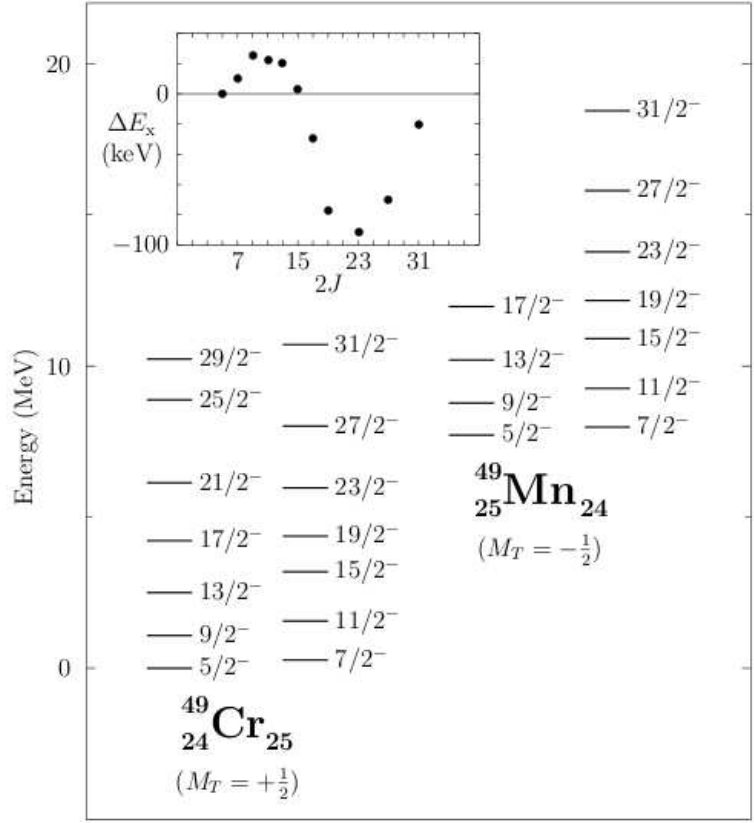}
\caption{Energy spectra of the mirror nuclei
$^{49}$Cr and $^{49}$Mn
relative to the ground state of the first nucleus.
Levels are labeled
by their angular momentum and parity $J^\pi$.
The inset shows the difference in excitation energy
$\Delta E_{\rm x}\equiv
E_{\rm x}(^{49}{\rm Cr};J)-E_{\rm x}(^{49}{\rm Mn};J)$
as a function of $2J$.}
\label{f_mass49}
\end{figure}

The assumption of isospin symmetry is too strong
and should be relaxed to one of dynamical symmetry.
One cannot expect that isobaric analogue states
have the same {\em absolute} energy
but one can expect them to have, to a good approximation,
the same {\em relative} energies.
As a result, for example,
the excitation spectra of two mirror nuclei
should be identical
although the binding energy of their ground states differs.
(Mirror nuclei have the same total number of nucleons
and the number of neutrons in one of them
equals the number of protons in the other.)
This relation has been observed in many cases.
An example where the idea has been tested
to high angular momentum,
is shown in Fig.~\ref{f_mass49}~\cite{Leary97}.
The ground-state energies of the two nuclei
of the $T={\frac 1 2}$ isospin doublet
($^{49}{\rm Cr}$ with $M_T=+{\frac 1 2}$
and $^{49}{\rm Mn}$ with $M_T=-{\frac 1 2}$)
are shifted with respect to each other
but the energies relative to the ground state
are indeed very similar.
Nevertheless, the spectra are not identical
as is clear from the inset in Fig.~\ref{f_mass49}
where the difference in excitation energy
is plotted as a function of the angular momentum $J$.
The deviations from zero signal
a breakdown of the dynamical-symmetry approximation
and, specifically, reveal subtle differences
in alignment properties of the neutrons and protons
in the two mirror nuclei~\cite{Bentley07}.

\begin{figure}
\centering
\includegraphics[width=10cm]{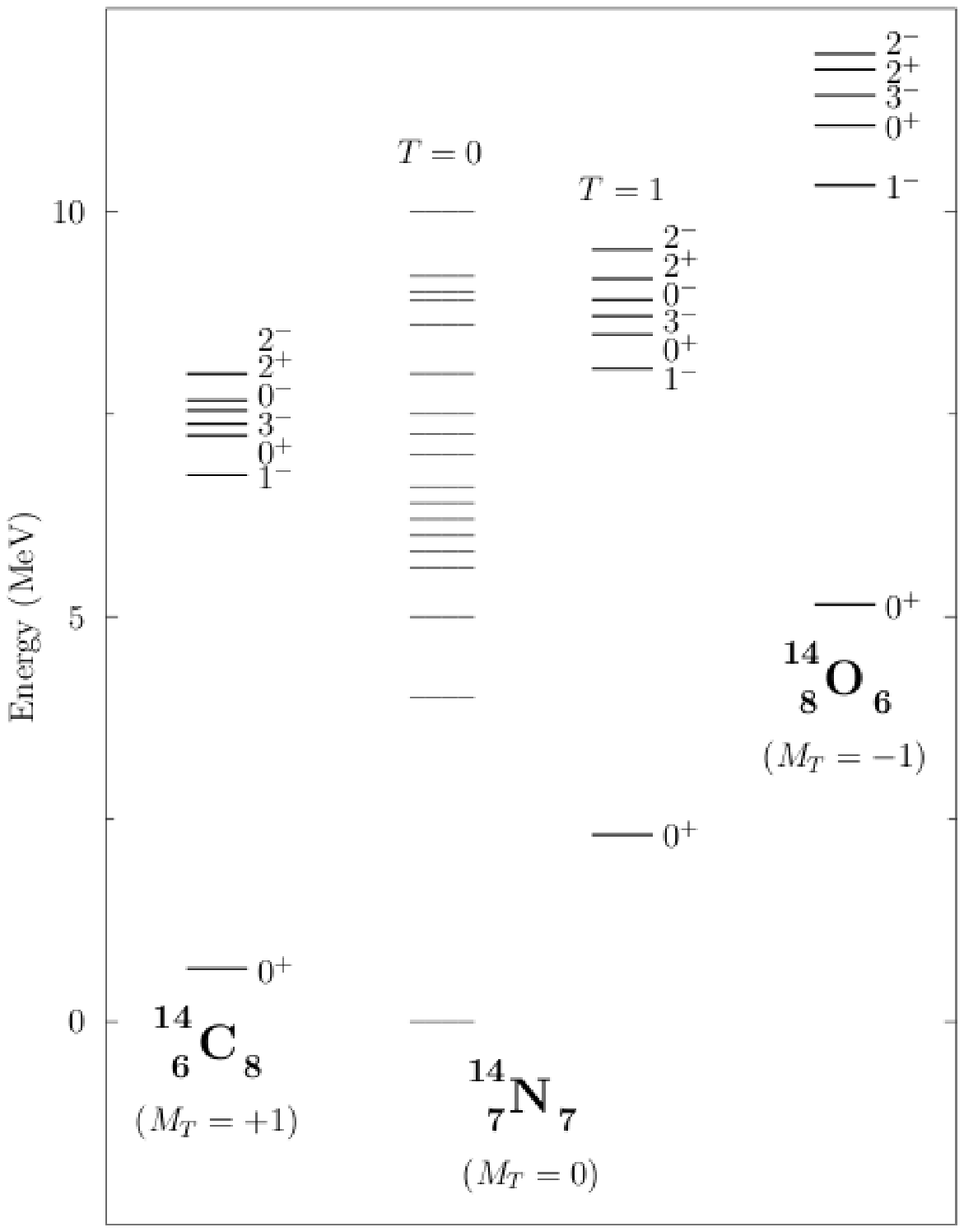}
\caption{Energy spectra of the nuclei
$^{14}$C, $^{14}$N and $^{14}$O
relative to the ground state of the middle nucleus.
States with isospin $T=1$ are drawn in thick lines.
In the self-conjugate nucleus $^{14}$N
there exist also states with isospin $T=0$
which are drawn in thin lines.
Levels with $T=1$ are labeled
by their angular momentum and parity $J^\pi$.}
\label{f_mass14}
\end{figure}

The equality of excitation spectra of mirror nuclei
is sometimes referred to as mirror symmetry.
It should be emphasized that mirror symmetry
is but a particular manifestation of isospin symmetry
which implies a wider relationship between properties of nuclei
as illustrated with the example in Fig.~\ref{f_mass14}.
The nuclei shown contain $A=14$ nucleons
but differ by their numbers of neutrons and protons,
$(N,Z)=(8,6)$, (7,7) and (6,8).
This corresponds to eigenvalues of $T_z$
given by $M_T=(N-Z)/2=+1,0,-1$
and, consequently,
the isospin of all states in $^{14}$C and $^{14}$O
must be $T=1$ or higher.
As a consequence of mirror symmetry,
the low-energy spectra of both nuclei should be identical.
The $T=1$ analogue states should also occur in $^{14}$N, however.
This nucleus has $M_T=0$
but this does not preclude the existence of $T=1$ states.
In fact, isospin symmetry {\em requires}
that such states be present somewhere in the spectrum of $^{14}$N.
Figure~\ref{f_mass14} illustrates
that the isobaric analogue levels of those in $^{14}$C and $^{14}$O
are indeed found in $^{14}$N.

\begin{figure}
\centering
\includegraphics[width=11.5cm]{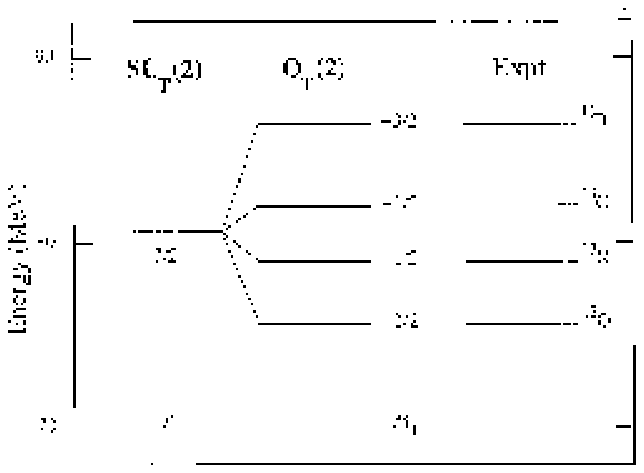}
\caption{
Binding energies of the $T={\frac 3 2}$ isobaric analogue states
with $J^\pi={\frac 1 2}^-$ in $^{13}$B, $^{13}$C, $^{13}$N and $^{13}$O.
The column on the left is obtained
for an exact ${\rm SU}_T(2)$ isospin symmetry,
which predicts states with different $M_T$ to be degenerate.
The middle column is obtained
with the IMME with $\kappa_0=80.59$, $\kappa_1=-2.96$
and $\kappa_2=-0.26$, in MeV.}
\label{f_mass13}
\end{figure}

For $T\geq{\frac 3 2}$ it is possible to test the IMME
since the parameters $\kappa_i$
can be fixed from the isobaric analogue states in three nuclei
and a prediction follows for the fourth member of the multiplet.
As an example consider the $T={\frac 3 2}$ multiplet
consisting of isobaric analogue states in
$^{13}$B, $^{13}$C, $^{13}$N and $^{13}$O.
Figure~\ref{f_mass13} shows the binding energies
of the nuclei $^{13}$B and $^{13}$O,
both of which have $T=|M_T|={\frac 3 2}$ in their ground state.
The isobaric analogue states in $^{13}$C and $^{13}$N
are $J^\pi={\frac 1 2}^-$ states
at excitation energies of 15.11 and 15.07~MeV, respectively;
these energies are substracted
from the ground-state binding energies of $^{13}$C and $^{13}$N
to give the energies plotted in Fig.~\ref{f_mass13}.
In this example the energy splitting due to the Coulomb interaction
is well accounted for by the IMME,
which is perhaps not surprising since four data points
are fitted with three parameters.
The quality of fits such as the one in Fig.~\ref{f_mass13}
is, however, not the most important aspect of dynamical symmetries,
but rather the existence of good quantum numbers
(isospin $T$ in this case).

\subsection{Isospin selection rules}
\label{ss_sel}
The most important consequence of a symmetry,
which remains valid
under the process of a dynamical symmetry breaking,
is the existence of conserved (or `good') quantum numbers.
Frequently, these quantum numbers
give rise to selection rules in radiative transition
or particle-transfer processes.
The measurement of transition or transfer probabilities
is thus the method to establish the goodness of labels
needed to characterize a quantum state
and this in turn indicates
to what extent a given (dynamical) symmetry is valid.

The link between symmetries and selection rules
can be given a precise quantitative formulation
via the (generalized) Wigner--Eckart theorem.
Suppose the calculation is required
of a transition or transfer matrix element
between an initial state $|\Gamma_{\rm i}\gamma_{\rm i}\rangle$
and a final state $|\Gamma_{\rm f}\gamma_{\rm f}\rangle$,
where the labeling of Subsect.~\ref{ss_deg} is adopted.
To compute the matrix element,
it is first necessary
to determine the tensor character of the operator
associated with the transition or transfer
by formally writing the operator as
$\sum_{\Gamma\gamma}a_{\Gamma\gamma}T^\Gamma_\gamma$
where $a_{\Gamma\gamma}$ are coefficients.
Each piece $T^\Gamma_\gamma$
can now be dealt with separately
through the generalized Wigner--Eckart theorem.
The essential point is that all dependence
on the quantum numbers associated with the subalgebra ${\rm G}_2$
is contained in a generalized coupling coefficient.
In addition, selection rules now follow
from the multiplication rules
for irreducible representations of the algebra ${\rm G}_1$:
if $\Gamma_{\rm f}$ is not contained
in the product $\Gamma_{\rm i}\times\Gamma$,
the generalized coupling coefficient is zero
and the matrix element of $T^\Gamma_\gamma$ vanishes.

A well-known example of the idea of selection rules
concerns electric dipole transitions
in self-conjugate nuclei~\cite{Trainor52,Radicati52},
that is, nuclei with an equal number
of neutrons and protons ($N=Z$).
The E1 operator is,
in lowest order of the long-wave approximation,
given by
\begin{equation}
T_\mu({\rm E1})=\sum_{k=1}^A e_kr_\mu(k).
\end{equation}
Since the charge $e_k$ of the $k^{\rm th}$ nucleon
is zero for a neutron and $e$ for a proton,
the E1 operator can be rewritten as
\begin{equation}
T_\mu({\rm E1})=
{\frac e 2}\sum_{k=1}^A[1-2t_z(k)]r_\mu(k)=
{\frac e 2}\left[R_\mu-2\sum_{k=1}^At_z(k)r_\mu(k)\right],
\end{equation}
where $2t_z$
gives $+1$ for a neutron and $-1$ for a proton.
The first term $R_\mu$ in the E1 operator
is the centre-of-mass coordinate of the total nucleus
and does not contribute to an internal E1 transition.
The conclusion is
that the electric dipole operator
is, in lowest order of the long-wave approximation,
of pure isovector character.
The application of the Wigner--Eckart theorem
in isospin space gives
\begin{equation}
\langle\eta_{\rm f}T_{\rm f}M_{T_{\rm f}}|
T^{(1)}_0
|\eta_{\rm i}T_{\rm i}M_{T_{\rm i}}\rangle=
\langle T_{\rm i}M_{T_{\rm i}}\;10|
T_{\rm f}M_{T_{\rm f}}\rangle
\langle\eta_{\rm f}T_{\rm f}\|
T^{(1)}
\|\eta_{\rm i}T_{\rm i}\rangle,
\end{equation}
where the coupling coefficient
is associated with ${\rm SU}(2)\supset{\rm SO}(2)$.
Self-conjugate nuclei have $M_{T_{\rm i}}=M_{T_{\rm f}}=0$
and exhibit as a consequence a simple selection rule:
E1 transitions are forbidden
between levels with the same isospin $T_{\rm i}=T_{\rm f}=T$
because of the vanishing Clebsch--Gordan coefficient,
$\langle T0\;10|T0\rangle=0$.

\begin{figure}
\centering
\includegraphics[width=10cm]{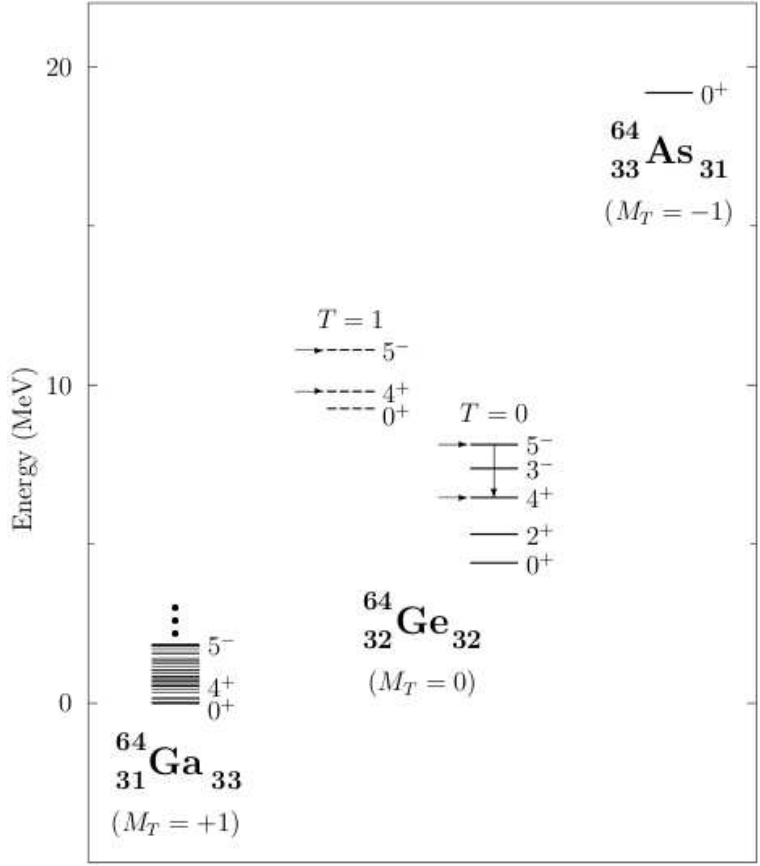}
\caption{Energy spectra of the nuclei
in the $A=64$ isospin triplet
$^{64}$Ga, $^{64}$Ge and $^{64}$As
relative to the ground state of the first nucleus.
The observed $5^-\rightarrow4^+$ E1 transition
between $T=0$ states in $^{64}$Ge is explained
through mixing with the $T=1$ states,
indicated by the arrows.
The levels in broken lines
are inferred from the isospin analogue levels
in $^{64}$Ga.}
\label{f_mass64}
\end{figure}

This selection rule has been verified
to hold approximately
in light self-conjugate nuclei~\cite{Freeman66}.
Deviations occur
because of higher-order terms in the E1 operator
but also, and more importantly,
because isospin is not an exactly conserved quantum number.
Isospin mixing can be estimated
in a variety of nuclear models.
They all show that the mixing
({\it i.e.}, the non-dynamical breaking of isospin symmetry)
is maximal in $N=Z$ nuclei.
Isospin mixing effects,
caused mainly by the Coulomb interaction,
should thus be looked for in heavy $N=Z$ nuclei
where they are largest.
Such nuclei are created and accelerated for study
at radioactive-ion beam facilities.
The spectrum of an $N=Z$ nucleus studied in this respect, $^{64}$Ge,
is shown in Fig.~\ref{f_mass64}.
The crucial transition is the E1
between the $5^-$ and $4^+$ levels
(indicated by the down arrow)
which should be strictly forbidden
if the isospin dynamical symmetry were exact.
A small $B({\rm E1};5^-\rightarrow4^+)$ value
is measured nevertheless
and this is explained through the mixing
with higher-lying $5^-$ and $4^+$ levels
in $^{64}$Ge with $T=1$,
which are not observed but inferred
from their isospin analogue states in $^{64}$Ga.
Although an estimate of the isospin mixing
can be made in this way,
the procedure is difficult
as it requires the measurement of the lifetime,
the $\delta({\rm E2}/{\rm M1})$ mixing ratio
and the relative intensities of the transitions
de-exciting the $5^-$ level~\cite{Farnea03}.
Given these uncertainties,
a reliable measurement of isospin admixtures in nuclei,
as a function of $N$ and $Z$,
is still a declared goal
of the current experimental efforts
with radioactive-ion beams.

\section{The nuclear shell model}
\label{s_shell}
The structure of the atomic nucleus is determined,
in first approximation, by the nuclear mean field,
the average potential felt by a nucleon
through the interactions exerted by all others.
This average potential is responsible for the shell structure of the nucleus
because the energy spectrum of a particle moving in this mean field
shows regions with many levels and others with few.
A second important ingredient that determines the structure of nuclei
(and generally of many-body quantum systems) is the Pauli principle.
Consequently, the nucleus can be viewed as an onion-like construction,
with shells determined by the mean-field potential
that are being filled in accordance with the Pauli principle.
For a description that goes beyond this most basic level,
the residual interaction between nucleons must be taken into account
and what usually matters most for nuclear structure at low energies
is the residual interaction between nucleons in the valence or outer shell.
This interaction depends in a complex fashion
on the numbers of valence neutrons and protons,
and on the valence orbits available to them.

No review is given here of the nuclear shell model
which has been the subject of several comprehensive
monographs~\cite{Shalit63,Bohr69,Lawson80,Heyde90,Talmi93}.
Instead, after an introductory subsection,
describing the model's essential features and assumptions,
emphasis is laid on its symmetry structure.
It turns out that the two most important correlations in nuclei,
of the pairing and of the quadrupole type, respectively,
can be analyzed with symmetry techniques.

\subsection{The model}
\label{ss_shell}
In a non-relativistic approximation,
the wavefunction of any quantum-mechanical state
of a nucleus with $A$ nucleons
satisfies the Schr\"odinger equation
\begin{equation}
H
\Psi(\xi_1,\xi_2,\dots,\xi_A)=
E\Psi(\xi_1,\xi_2,\dots,\xi_A),
\label{e_schrod}
\end{equation}
with the Hamiltonian
\begin{equation}
H=
\sum_{k=1}^A {{p^2_k}\over{2m_k}}
+\sum_{k<l}^AW_2(\xi_k,\xi_l)
+\sum_{k<l<m}^AW_3(\xi_k,\xi_l,\xi_m)
+\cdots.
\label{e_hamsm}
\end{equation}
The notation $\xi_k$ is used
to denote all coordinates of nucleon $k$,
not only its position vector $\bar r_k$
but also its spin $\bar s_k$
and its isospin $\bar t_k$,
$\xi_k\equiv\{\bar r_k,\bar s_k,\bar t_k\}$.
The term $p^2_k/2m_k$ is the kinetic energy of nucleon $k$
and acts on a single nucleon only.
The operator $W_i(\xi_k,\xi_l,\xi_m,\dots)$ is
an $i$-body interaction between the nucleons $k,l,m,\dots$,
and, as such, acts on $i$ nucleons simultaneously.
Since neutron and proton
are not elementary particles,
it is not {\it a priori} clear that the interaction
should be of two-body nature.
Nevertheless, for a presentation
of the elementary nuclear shell model,
it can be assumed that the nature
between the nucleons is two-body, $W_{i>2}=0$,
as will be done in the subsequent discussion.

Under the assumption of at most two-body interactions,
one can rewrite (\ref{e_hamsm}) as
\begin{equation}
H=
\sum_{k=1}^A
\left({{p^2_k}\over{2m_k}}+V(\xi_k)\right)
+\left(
\sum_{k<l}^AW_2(\xi_k,\xi_l)
-\sum_{k=1}^AV(\xi_k)
\right).
\label{e_hamsm2}
\end{equation}
The idea is now to choose $V(\xi_k)$
such that the effect of the residual interaction,
that is, the second term in~(\ref{e_hamsm2}), is minimized.
The independent-particle shell model is obtained
by neglecting the residual interaction altogether,
\begin{equation}
H_{\rm ip}=
\sum_{k=1}^A
\left({{p^2_k}\over{2m_{\rm n}}}+V(\xi_k)\right),
\label{e_hamip}
\end{equation}
where it is also assumed that all nucleons
have the same mass $m_{\rm n}$.
The physical interpretation of the approximation~(\ref{e_hamip})
is that each nucleon moves independently
in a mean-field potential $V(\xi)$
which represents the average interaction
with all other nucleons in the nucleus.

The eigenproblem associated with the Hamiltonian~(\ref{e_hamip})
is much easier to solve than the original problem~(\ref{e_schrod})
because it can be reduced to a one-particle eigenequation.
Its solution proceeds as follows.
First, one solves the Schr\"odinger equation
of a particle in a potential $V(\xi)$,
that is,
one finds the eigenfunctions $\phi_i(\xi)$ satisfying
\begin{equation}
\left({{p^2}\over{2m_{\rm n}}}+V(\xi)\right)\phi_i(\xi)=
E_i\phi_i(\xi),
\label{e_onepart}
\end{equation}
where $i$ labels the different eigensolutions.
The exact form of the eigenfunctions $\phi_i(\xi)$
depends on the potential $V(\xi)$.
For simple potentials ({\it e.g.}, the harmonic oscillator)
the eigenfunctions can be found in analytic form
in terms of standard mathematical functions;
for more complicated potentials ({\it e.g.}, Woods--Saxon)
$\phi_i(\xi)$ must be determined numerically.
For all `reasonable' potentials $V(\xi)$
the solutions of~(\ref{e_onepart}) can be obtained,
albeit in most cases only in numerical form.

The solution of the {\em many-body} Hamiltonian $H_{\rm ip}$
is immediately obtained due to its separability,
\begin{equation}
\Phi_{i_1i_2\dots i_A}(\xi_1,\xi_2,\dots,\xi_A)=
\prod_{k=1}^A
\phi_{i_k}(\xi_k).
\label{e_onesol}
\end{equation}
Although this is a genuine, mathematical eigensolution
of the Hamiltonian~(\ref{e_hamip}),
it is not antisymmetric under the exchange of particles
as is required by the Pauli principle.
The solution~(\ref{e_onesol}) must thus be antisymmetrized.
For $A=2$ particles the antisymmetrization procedure yields
\begin{eqnarray}
\Psi_{i_1i_2}(\xi_1,\xi_2)&=&
\sqrt{1\over2}
[\phi_{i_1}(\xi_1)\phi_{i_2}(\xi_2)
-\phi_{i_1}(\xi_2)\phi_{i_2}(\xi_1)]
\nonumber\\&=&
\sqrt{1\over2}
\left|\begin{array}{cc}
\phi_{i_1}(\xi_1)&\phi_{i_1}(\xi_2)\\
\phi_{i_2}(\xi_1)&\phi_{i_2}(\xi_2)
\end{array}\right|.
\end{eqnarray}
In the $A$-particle case,
antisymmetrization leads to the replacement of the wave function
$\Phi_{i_1i_2\dots i_A}(\xi_1,\xi_2,\dots,\xi_A)$
by a Slater determinant of the form
\begin{equation}
\Psi_{i_1i_2\dots i_A}(\xi_1,\xi_2,\dots,\xi_A)=
{1\over\sqrt{A!}}
\left|\begin{array}{cccc}
\phi_{i_1}(\xi_1)&\phi_{i_1}(\xi_2)&\cdots&\phi_{i_1}(\xi_A)\\
\phi_{i_2}(\xi_1)&\phi_{i_2}(\xi_2)&\cdots&\phi_{i_2}(\xi_A)\\
\vdots&\vdots&\ddots&\vdots\\
\phi_{i_A}(\xi_1)&\phi_{i_A}(\xi_2)&\cdots&\phi_{i_A}(\xi_A)
\end{array}\right|.
\label{e_slater}
\end{equation}
This is the solution of the Schr\"odinger equation
associated with the Hamiltonian~(\ref{e_hamip})
that takes account of the Pauli principle.

The following question now arises.
How should one choose the potential $V(\xi)$
introduced in~(\ref{e_hamsm2})?
This choice can be made at several levels of refinement.
Ideally one wants to minimize
the expectation value of $H$ in the ground state, that is,
solve the variational equation
\begin{equation}
\delta\int
\Psi^*(\xi_1,\xi_2,\dots,\xi_A)
H
\Psi(\xi_1,\xi_2,\dots,\xi_A)
d\xi_1d\xi_2\dots d\xi_A=0.
\label{e_var}
\end{equation}
If, in this variational approach,
the wave function $\Psi(\xi_1,\xi_2,\dots,\xi_A)$
is allowed to vary freely,
the solution of~(\ref{e_var}) is equivalent
to the ground-state solution
of the Schr\"odinger equation~(\ref{e_schrod}).
Obviously, one needs to set more modest goals
to arrive at a solvable problem!
One way to do so
is to restrict $\Psi(\xi_1,\xi_2,\dots,\xi_A)$
in~(\ref{e_var}) to the form of a Slater determinant,
in other words, to minimize the ground-state energy
by varying the potential $V(\xi)$
that defines the single-particle wave functions
$\phi_{i_1},\phi_{i_2},\dots,\phi_{i_A}$ in~(\ref{e_slater}).
This is known as the Hartree--Fock method.
One determines the form of the potential $V(\xi)$
by requiring the expectation value
of the {\em complete} Hamiltonian~(\ref{e_hamsm})
in the state~(\ref{e_slater}) to be minimal.

The ground-state energy determined in Hartree--Fock theory
is not the correct one;
nevertheless, it is the best procedure at hand
to construct an independent-particle model.
Its disadvantage is that it can be computationally rather involved.
Therefore, often the following simpler approach is preferred.
One proposes a phenomenological form
of the potential $V(\xi)$,
such that the Schr\"odinger equation
associated with $H_{\rm ip}$ in~(\ref{e_hamip})
is analytically solvable.
The potential which best mimics
the nuclear mean-field potential
and which can be solved exactly,
is the harmonic-oscillator potential
\begin{equation}
V(\xi)\equiv V(r)=
{\frac 1 2}m_{\rm n}\omega^2r^2.
\label{e_hopot}
\end{equation}
The eigensolutions of the Schr\"odinger equation
of a harmonic oscillator in three dimensions can be written as
\begin{equation}
\phi_{n\ell m_\ell}(r,\theta,\varphi)
=R_{n\ell}(r)Y_{\ell m_\ell}(\theta,\varphi),
\label{e_howave}
\end{equation}
where $R_{n\ell}(r)$ are radial wave functions
appropriate for the harmonic oscillator
and $Y_{\ell m_\ell}(\theta,\varphi)$ are spherical harmonics,
already introduced in Eq.~(\ref{e_hywave}).
The index $i$,
used previously to characterize single-particle eigenfunctions,
is replaced now by the full set
of quantum numbers $n$, $\ell$ and $m_\ell$.
The quantized energy spectrum is given by
\begin{equation}
E(n,\ell)=\left(2n+\ell+{\textstyle{3\over2}}\right)\hbar\omega,
\label{e_hoener}
\end{equation}
in terms of the radial quantum number $n$
which has the allowed values $0,1,2,\dots$
and gives the number of nodes
[values of $r$ for which $R_{n\ell}(r)=0$
excluding those at $r=0$ and $r=\infty$].
Because of the factor $r^\ell$ in the radial part,
the wave function always vanishes at $r=0$
except for $\ell=0$ ($s$ state).
The energy $E(n,\ell)$ is independent of $m_\ell$,
the projection of the orbital angular momentum along the $z$ axis,
as should be for a rotationally invariant Hamiltonian.
In addition, $E(n,\ell)$ is only dependent on the sum $2n+\ell$.
Introducing $N=2n+\ell$, one can rewrite~(\ref{e_hoener}) as
\begin{equation}
E(N)=\left(N+{\textstyle{3\over2}}\right)\hbar\omega,
\end{equation}
which shows that $N$ can be interpreted
as the number of oscillator quanta,
the term ${3\over2}\hbar\omega$ being accounted for
by the zero-point motion of an oscillator in three dimensions;
$N$ is called the major oscillator quantum number.
The allowed values of the orbital angular momentum are
(because $\ell=N-2n$ and $n=0,1,\dots$)
\begin{equation}
\ell=N,N-2,\dots,0\;{\rm or}\;1.
\end{equation}
This completely determines the eigenspectrum
of a spinless particle in a harmonic-oscillator potential.

The $2\ell+1$ eigensolutions
with the same radial quantum number $n$
and the same orbital angular momentum $\ell$
but different $z$ projections $m_\ell$ are degenerate in energy.
This degeneracy arises
because the harmonic-oscillator Hamiltonian
is rotationally invariant.
There exists an {\em additional} degeneracy,
namely the one for levels with the same $2n+\ell$.
As in the case of the spectrum of the hydrogen atom,
discussed in Sect.~\ref{s_hydrogen},
this additional degeneracy is also associated
with a symmetry of the Hamiltonian
which is identified in this case as U(3)~\cite{Moshinsky69}.
The U(3) transformations
are more general than rotations in three dimensions
[{\it i.e.}, U(3) contains SO(3)]
and U(3) invariance can be understood intuitively
as a consequence of the equivalence between the excitation of quanta
in the $x$, $y$ and $z$ directions.
The degeneracies of the harmonic-oscillator energy levels
do not occur for a Woods--Saxon potential.
In general one finds for a Woods--Saxon potential
that, of the orbits with the same major oscillator quantum number $N$,
those with high $\ell$ are more strongly bound
than those with low orbital angular momentum.

An important quantity appearing in the harmonic-oscillator model
is the elementary quantum of excitation $\hbar\omega$.
By relating the radius of the nucleus, $R$,
to the number of nucleons, $A$,
and subsequently deriving a relationship
between $R$, $A$ and the oscillator length $b$,
one finds the expression~\cite{Bohr69}
\begin{equation}
b\approx1.00 A^{1/6}\;{\rm fm},
\label{ip16}
\end{equation}
and, since $b=\sqrt{\hbar/m_{\rm n}\omega}$,
\begin{equation}
\hbar\omega\approx41A^{-1/3}\;{\rm MeV}.
\end{equation}

The solutions $\phi_{n\ell m_\ell}(r,\theta,\varphi)$
contain the dependence on the spatial coordinates only
and not on the intrinsic spin of the particle.
Since the intrinsic spin
does not appear in the potential~(\ref{e_hopot}),
the wave functions are simply given by
\begin{equation}
\phi_{n\ell m_\ell}(r,\theta,\varphi)
\chi_{sm_s},
\label{e_spin}
\end{equation}
where $\chi_{sm_s}$ are spinors
for particles with intrinsic spin $s={\frac 1 2}$.
The energies are independent of $m_s$
and are still given by~(\ref{e_hoener}).
The eigenstates (\ref{e_spin})
do not have good {\em total} angular momentum,
that is, they are not eigenstates of $j^2$
where $\bar j$ results from the coupling
of the orbital angular momentum $\bar\ell$ and the spin $\bar s$ of the nucleon.
States of good angular momentum are constructed
from~(\ref{e_spin}) with the help of Clebsch--Gordan coefficients,
\begin{equation}
\phi_{n\ell jm_j}(r,\theta,\varphi)=
\sum_{m_\ell m_s}
(\ell m_\ell\:sm_s|jm_j)
\phi_{n\ell m_\ell}(r,\theta,\varphi)
\chi_{sm_s}.
\end{equation}
Again, this state has the same energy eigenvalue~(\ref{e_hoener})
since all states appearing in the sum are degenerate.

If the spin degeneracy
of the quantum numbers $(n\ell jm_j)$ is taken into account,
stable shell gaps are obtained
at the nucleon numbers 2, 8, 20, 40, 70, 112,\dots.
These are the magic numbers of the harmonic oscillator.

\begin{figure}
\centering
\includegraphics[width=12cm]{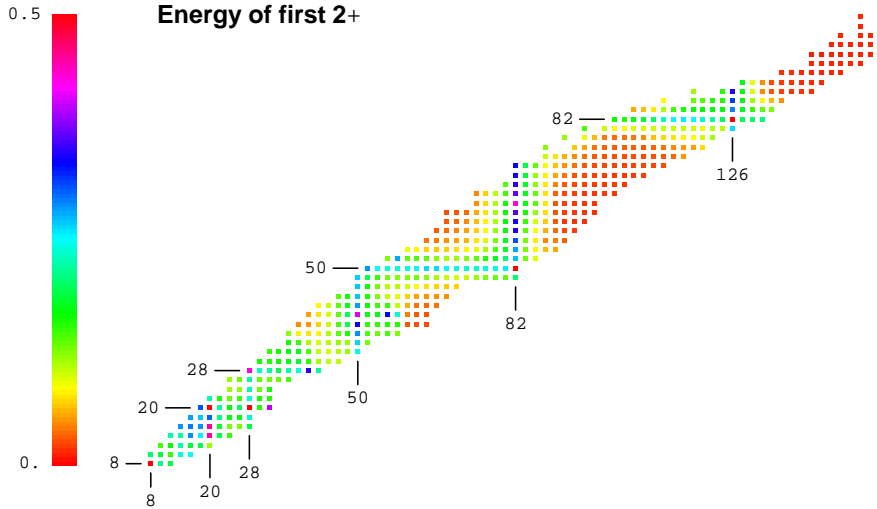}
\caption{The energy of the first-excited $2^+$ state
in all even--even nuclei with $N,Z\geq8$
(where known experimentally)
plotted as a function of neutron number $N$ along the $x$ axis
and proton number $Z$ along the $y$ axis.
The excitation energy is multiplied by $A^{1/3}$
and subsequently normalized to 1 for $^{208}$Pb
where this quantity is highest.
The value of $E_{\rm x}(2_1^+)A^{1/3}$
is indicated by the scale shown on the left.
To improve the resolution of the plot,
the scale only covers part of the range from 0 to 0.5
since only a few doubly magic nuclei
($^{16}$O, $^{40,48}$Ca, $^{132}$Sn and $^{208}$Pb)
have values greater than 0.5.}
\label{f_twoplus}
\end{figure}

The existence of nuclear shell structure
can be demonstrated in a variety of ways.
The most direct way is by measuring
the ease with which a nucleus can be excited.
If it has a closed shell structure,
one expects it to be rather stable and difficult to excite.
This should be particularly so for nuclei that are doubly magic,
that is, nuclei with a closed-shell configuration
for neutrons {\em and} protons.
The principle is illustrated in Fig.~\ref{f_twoplus}.
The figure shows the energy $E_{\rm x}(2^+_1)$
of the first-excited $2^+$ state
relative to the ground state for all even--even nuclei.
This energy is multiplied with $A^{1/3}$
and the result plotted on a normalized scale.
(The factor $A^{1/3}$ accounts
for the gradual decrease with mass number $A$
of the strength of the nuclear residual interaction
which leads a compression of the spectrum with $A$.)
Nuclei with particularly high values of $E_{\rm x}(2^+_1)A^{1/3}$
are $^{16}$O ($N=Z=8$),
$^{40}$Ca ($N=Z=20$),
$^{48}$Ca ($N=28$, $Z=20$),
$^{132}$Sn ($N=82$, $Z=50$) and
$^{208}$Pb ($N=126$, $Z=82$).
Figure~\ref{f_twoplus} establishes the stability properties
of the isotopes and/or isotones
with $N,Z=8$, 20, 28, 50, 82 and 126.

How to explain the differences
between the observed magic numbers (2, 8, 20, 28, 50, 82 and 126)
and those of the harmonic oscillator?
The observed ones can be reproduced
in an independent-particle model
if to the harmonic-oscillator Hamiltonian $H_{\rm ho}$
a spin--orbit as well as an orbit--orbit term
is added of the form
\begin{equation}
V_{\rm so}=\zeta_{\rm so}(r)\bar\ell\cdot\bar s,
\qquad
V_{\rm oo}=\zeta_{\rm oo}(r)\bar\ell\cdot\bar\ell.
\end{equation}
The eigenvalue problem
associated with the Hamiltonian
$H_{\rm ho}+V_{\rm so}+V_{\rm oo}$
is not, in general, analytically solvable
but the dominant characteristics
can be found from the expectation values
\begin{equation}
\langle n\ell jm_j|V_{\rm so}|n\ell jm_j\rangle=
{\textstyle{1\over2}}
\langle\zeta_{\rm so}(r)\rangle_{n\ell}
\left[j(j+1)-\ell(\ell+1)-{\textstyle{3\over4}}\right],
\end{equation}
and
\begin{equation}
\langle n\ell jm_j|V_{\rm oo}|n\ell jm_j\rangle=
\ell(\ell+1)\langle\zeta_{\rm oo}(r)\rangle_{n\ell},
\end{equation}
with radial integrals defined as
\begin{equation}
\langle\zeta(r)\rangle_{n\ell}
=\int_0^{+\infty}
\zeta(r)R_{n\ell}(r)R_{n\ell}(r)r^2\;dr.
\end{equation}
Consequently, the degeneracy of the single-particle levels
within one major oscillator shell is lifted.
Empirically, one finds that the radial integrals
approximately satisfy the relations~\cite{Bohr69}
\begin{equation}
\langle\zeta_{\rm so}(r)\rangle_{n\ell}
\approx-20A^{-2/3}\;{\rm MeV},
\qquad
\langle\zeta_{\rm oo}(r)\rangle_{n\ell}
\approx-0.1\;{\rm MeV}.
\end{equation}
The origin of the orbit--orbit coupling
can be understood from elementary arguments.
The corrections to the harmonic-oscillator potential
are repulsive for short and large distances
and attractive for intermediate distances.
These corrections therefore
favor large-$\ell$ over small-$\ell$ orbits.
The spin--orbit coupling has a relativistic origin.
An important feature is that the radial integral is negative,
reflecting the empirical finding
that states with parallel spin and orbital angular momentum
are pushed down in energy
while in the antiparallel case they are pushed up.

The summary of the preceding discussion is
that a simple approximation of the nuclear mean-field potential
consists of a three-dimensional harmonic oscillator
corrected with a spin--orbit and an orbit--orbit term.
If, in addition, a two-body residual interaction is included,
the many-body Hamiltonian that must be solved
acquires the following form:
\begin{equation}
H=
\sum_{k=1}^A\left(
\frac{p^2_k}{2m_{\rm n}}+
{\frac1 2}m_{\rm n}\omega^2r_k^2+
\zeta_{\rm oo}\,\bar\ell_k\cdot\bar\ell_k+
\zeta_{\rm so}\,\bar\ell_k\cdot\bar s_k\right)
+\sum_{k<l}V_{\rm res}(\xi_k,\xi_l),
\label{e_hamho}
\end{equation}
where the indices in the second sum
run over a {\em restricted} number of particles,
usually only the valence nucleons.
In spite of the severe simplifications
of the original many-body problem~(\ref{e_schrod}),
the solution of the Schr\"odinger equation
associated with the Hamiltonian~(\ref{e_hamho})
still represents a formidable problem
since the residual interaction must be diagonalized
in a basis of Slater determinants of the type~(\ref{e_slater}).
Even if one limits oneself to valence-shell excitations,
the dimension of the Hilbert space rapidly explodes
with increasing mass of the nucleus.
The $m$-scheme basis can be used to illustrate this.
Because of the antisymmetry of Slater determinants,
their number can be computed easily.
For $n$ neutrons and $z$ protons
distributed over $\Omega_n$ and $\Omega_z$ orbital states,
respectively, the dimension of the basis is
\begin{equation}
{\frac{\Omega_n!}{n!(\Omega_n-n)!}}
{\frac{\Omega_z!}{z!(\Omega_z-z)!}}.
\end{equation}
Application of this formula
to $^{28}$Si (in the $sd$ shell, $\Omega_n=\Omega_z=12,n=z=6$)
and to $^{78}$Y
(half-way between the magic numbers 28 and 50,
$\Omega_n=\Omega_z=22,n=z=11$)
illustrates the point
since it leads to dimensions of 8.5 $10^5$ and 5.0 $10^{11}$,
respectively.

Given the considerable effort it takes
to solve the nuclear many-body problem even only approximately,
any analytical solution of~(\ref{e_hamho})
that can be obtained through symmetry techniques
might be of considerable value.
In fact, the residual interaction
can approximately be written as pairing-plus-quadrupole,
\begin{equation}
V_{\rm res}(\xi_k,\xi_l)=
V_{\rm pairing}(\bar r_k,\bar r_l)+
V_{\rm quadrupole}(\bar r_k,\bar r_l),
\end{equation}
where the exact form of these interactions is defined below.
For particular values of the parameters in the mean field
and if either the pairing or the quadrupole residual interaction is dominant,
the eigenproblem~(\ref{e_hamho}) can be solved analytically.
Three situations arise, of which two are of interest:
\begin{enumerate}
\item
{\it No residual interaction.}
If $V_{\rm res}(\xi_k,\xi_l)=0$, the solution of (\ref{e_hamho})
reduces to a Slater determinant
built from harmonic-oscillator eigenstates.
\item
{\it Pairing interaction.}
If the residual interaction has a pure pairing character,
Racah's SU(2) model of pairing results.
This model is usually applied
in the $jj$-coupling limit of strong spin--orbit coupling.
\item
{\it Quadrupole interaction.}
If the residual interaction has a pure quadrupole character,
Elliott's SU(3) model of rotation results.
This model requires an $LS$-coupling scheme
which occurs in the absence of spin--orbit coupling.
\end{enumerate}

\begin{figure}
\centering
\includegraphics[width=8cm]{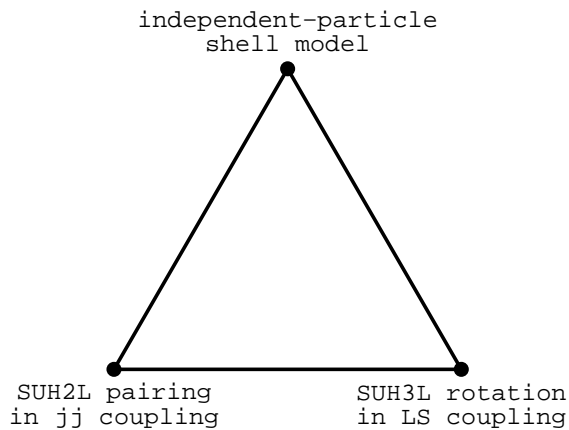}
\caption{Schematic representation
of the shell-model parameter space
with its three analytically solvable vertices.}
\label{f_trif}
\end{figure}
The situation is represented schematically in Fig.~\ref{f_trif}.
It should be emphasized that, in contrast to the top vertex,
the two bottom vertices, SU(2) and SU(3),
represent solutions of the nuclear Hamiltonian
which include genuine many-body correlations.
These two limits are thus of particular interest.
A brief summary of the pairing and quadrupole limits
of the nuclear shell model is given in the following subsections.
A more detailed review of the use of symmetries in the shell model
has been given elsewhere~\cite{Isacker99a}.

\subsection{Pairing correlations}
\label{ss_pair}
The pairing interaction is a reasonable first-order approximation
to the strong force between identical nucleons.
For nucleons in a single-$j$ shell
the interaction is defined by the matrix elements
\begin{equation}
\langle j^2;JM_J|V_{\rm pairing}|j^2;JM_J\rangle=
-g_0(2j+1)\delta_{J0},
\label{e_pair}
\end{equation}
where $j$ is the total (orbital+spin) angular momentum of a single nucleon
(hence $j$ is half-odd-integer),
$J$ results from the coupling of two $j$s
and $M_J$ is the projection of $J$ on the $z$ axis.
Furthermore, $g_0$ is the strength of the interaction
which is attractive in nuclei ($g>0$).

\begin{figure}
\centering
\includegraphics[width=11.5cm]{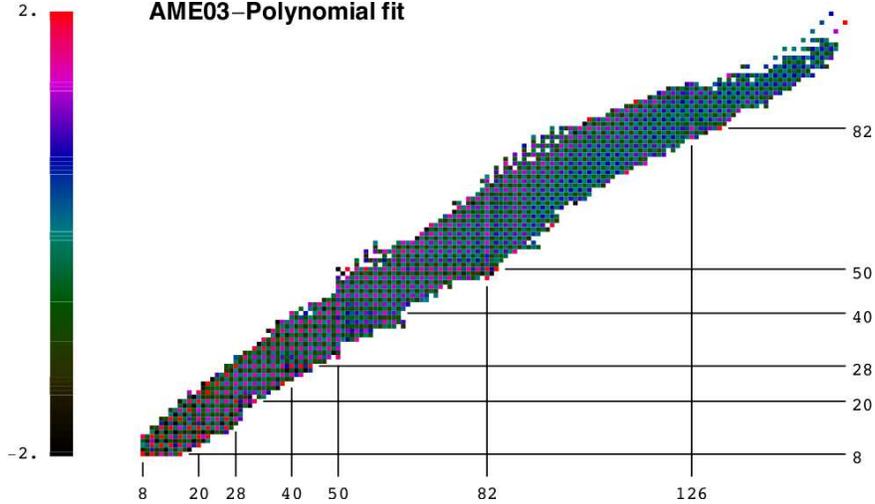}
\caption{The even--odd effect in nuclear binding energies
as evidence for pairing correlations in nuclei.
The difference between the experimental binding energies
from the atomic-mass compilation of 2003 (AME03)~\cite{Audi03}
and a smooth local fit to these data
is shown as a function of neutron number $N$ along the $x$ axis
and proton number $Z$ along the $y$ axis.
The local fit assumes a polynomial in $N$ and $Z$,
whose coefficients are determined
from about 50 masses in the neighborhood.}
\label{f_evenodd}
\end{figure}

Evidence for the pairing character
of the interaction between identical nucleons
can be obtained from simple arguments
as is illustrated in Fig.~\ref{f_evenodd}.
The figure shows the difference
between the experimental nuclear binding energies
and a smooth local fit to these data
as a function of neutron and proton numbers $N$ and $Z$.
The local fit assumes a polynomial in $N$ and $Z$,
whose coefficients are determined
to about 50 masses in the neighborhood.
The details of this fit are unimportant
for the present argument,
except for the fact that no difference is made
between even--even, odd-mass and odd--odd nuclei
which are all fitted with the same polynomial.
The figure clearly demonstrates the existence
of an even--odd effect in the observed binding energies
since even--even nuclei are systematically more bound
than found in the polynomial fit
while odd--odd nuclei are less bound.
The simplest interpretation of this empirical finding
is that there exists an attractive interaction
between two identical nucleons.

The pairing interaction
is less realistic than a short-range delta interaction
but has the advantage that the corresponding many-body problem
can be solved analytically.
Furthermore, its analysis is important
because it is at the basis of seniority~\cite{Racah43}
which has found fruitful application in nuclear physics
with considerable empirical evidence in semi-magic nuclei.

The results can be summarized as follows.
A state with $n$ identical particles
and diagonal in the pairing interaction,
is characterized---in addition to the angular momentum $J$
and its projection $M_J$---by a quantum number $\upsilon$.
The energy of this state is given by
\begin{equation}
E(n,\upsilon)=-{\frac 1 4}g_0(n-\upsilon)(2\Omega_j-n-\upsilon+2),
\label{e_paire}
\end{equation}
where $2\Omega_j=2j+1$.
The quantum number $\upsilon$
{\em counts the number of particles not coupled to $J=0$}.
Any state $|j^n\upsilon JM_J\rangle$ can be constructed from $|j^\upsilon\upsilon JM_J\rangle$
according to
\begin{equation}
|j^n\upsilon JM_J\rangle\propto
\left(S^j_+\right)^{(n-\upsilon)/2}|j^\upsilon\upsilon JM_J\rangle,
\label{e_pairw}
\end{equation}
where $S^j_+$ is an operator
which creates a pair of particles in the $j$ shell
with their angular momentum coupled to $J=0$.
In other words, $|j^\upsilon\upsilon JM_J\rangle$ acts as a {\em parent} state
for a whole class of states $|j^n\upsilon JM_J\rangle$
just by the action of the pair state $S^j_+$.
For this reason, $\upsilon$ is called seniority.

The above results remain valid if the $n$ identical particles
are distributed over several {\em degenerate} $j$ shells
by making the substitutions
$S^j_+\mapsto S_+\equiv\sum_jS^j_+$
and $\Omega_j\mapsto\Omega=\sum_j\Omega_j$.
In this form the pairing formalism
can be used to make several characteristic predictions:
a constant excitation energy (independent of $n$)
of the first-excited $2^+$ state in even--even isotopes,
the linear variation of two-nucleon separation energies as a function of $n$,
the odd--even staggering in nuclear binding energies,
the enhancement of two-nucleon transfer.

\begin{figure}
\centering
\includegraphics[width=10cm]{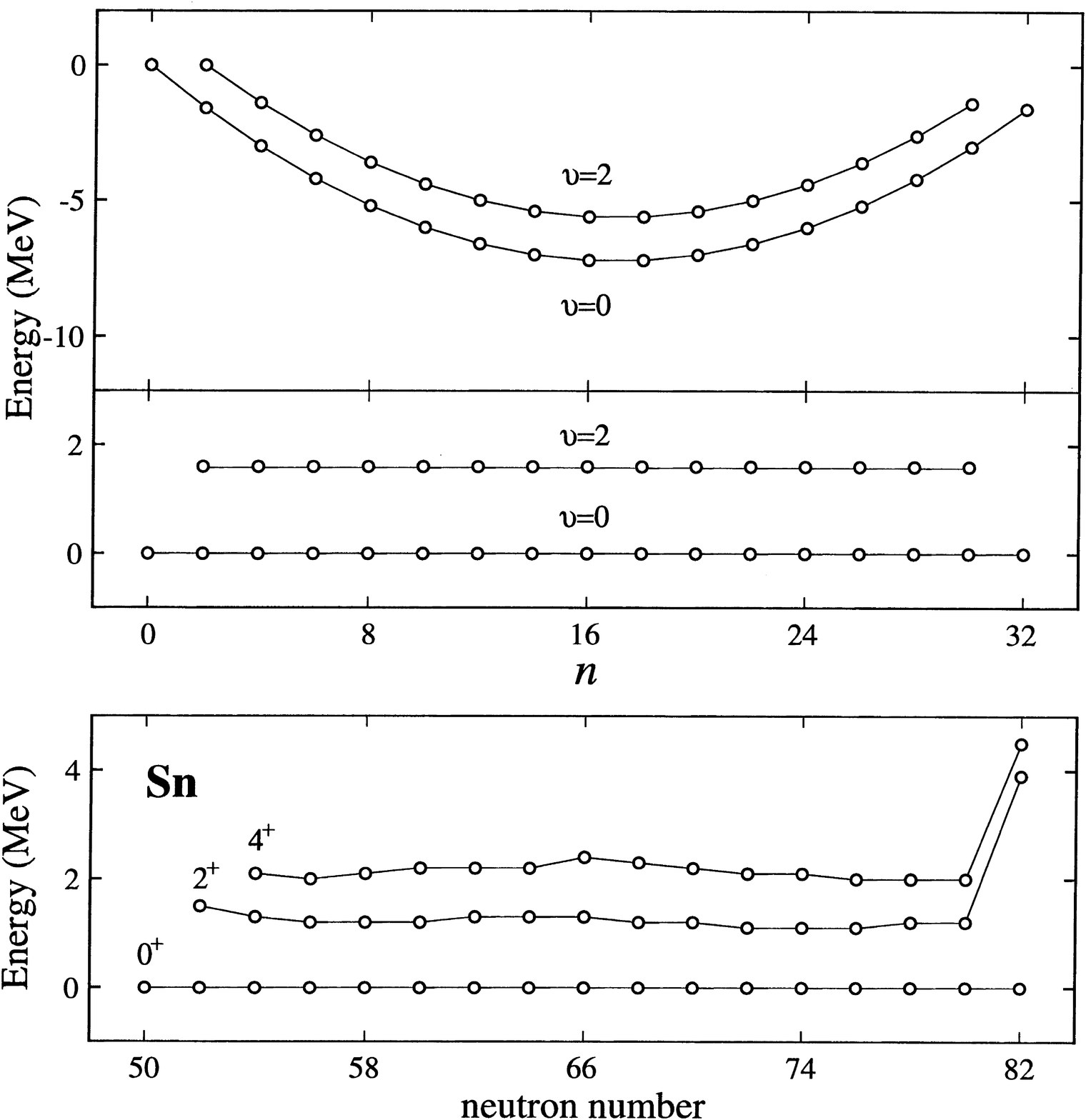}
\caption{The difference $E(n,2)-E(n,0)$
is a function of particle number $n$ (top)
and the corresponding observed excitation energies
$E_{\rm x}(2^+_1)\equiv E(2^+_1)-E(0^+_1)$
and $E_{\rm x}(4^+_1)\equiv E(4^+_1)-E(0^+_1)$
in the Sn isotopes.}
\label{f_sntwop}
\end{figure}
The first of these predictions is illustrated in Fig.~\ref{f_sntwop}.
The ground state of an even--even nucleus has $\upsilon=0$
and the lowest excited states have $\upsilon=2$.
An example of such $\upsilon=2$ states
are those in a two-nucleon $j^2$ configuration
with $J\neq0$, $J=2,4,\dots,2j-1$.
The energy difference between $\upsilon=2$ and $\upsilon=0$ states
is given by
\begin{equation}
E(n,2)-E(n,0)=g_0\Omega,
\end{equation}
and is independent of the number of valence nucleons.
This prediction is illustrated in Fig.~\ref{f_sntwop}
where it is compared with the excitation energies
of the $2^+_1$ and $4^+_1$ levels in the even--even Sn isotopes.

The discussion of pairing correlations in nuclei
traditionally has been inspired
by the treatment of superfluidity in condensed matter~\cite{Bardeen57,Bohr58}.
The superfluid phase in the latter systems
is characterized by the presence of a large number
of identical bosons in a single quantum state,
which is called the condensate.
In superconductors the bosons are pairs of electrons
with opposite momenta that form at the Fermi surface.
The character of the bosons in nuclei
can be understood by analyzing the ground state
of a pairing Hamiltonian.
For an even--even nucleus, according to~(\ref{e_pairw}),
it is given by
\begin{equation}
|j^n\upsilon=0,J=M=0\rangle\propto
\left(S_+\right)^{n/2}|{\rm o}\rangle.
\end{equation}
In nuclei the bosons
are thus pairs of valence nucleons with opposite angular momenta.

\subsection{Quadrupole correlations}
\label{ss_quad}
The second class of analytically solvable shell-model Hamiltonians
corresponds to the case of nucleons occupying
an entire shell of the harmonic oscillator
and interacting through a quadrupole force.
In this case the Hamiltonian is of the form
\begin{equation}
H=
\sum_{k=1}^A
\left({{p_k^2}\over{2m_{\rm n}}}+
{\frac 1 2}m_{\rm n}\omega^2r_k^2\right)-
g_2Q\cdot Q,
\label{e_su3ham}
\end{equation}
which contains a quadrupole operator
\begin{equation}
Q_\mu=
\sqrt{\frac 3 2}
\left[
\sum_{k=1}^A{\frac{1}{b^2}}
[\bar r_k\times\bar r_k]^{(2)}_\mu+
{\frac{b^2}{\hbar^2}}
\sum_{k=1}^A[\bar p_k\times\bar p_k]^{(2)}_\mu
\right].
\label{e_su3q}
\end{equation}
Note that $Q\cdot Q\equiv\sum_\mu Q_\mu Q_\mu$
contains one-body ($k=l$)
as well as two-body ($k\neq l$) terms.

The proof that the shell-model Hamiltonian~(\ref{e_su3ham})
is analytically solvable
was given by Elliott~\cite{Elliott58}.
The reasons for its solvability are that
the five components of the quadrupole operator~(\ref{e_su3q})
together with the three components
of the angular momentum vector
$\bar L=\sum_k(\bar r_k\wedge\bar p_k)$
form a closed algebra SU(3)
and, furthermore,  that these operators commute
with the harmonic-oscillator Hamiltonian
[{\it i.e.}, with the one-body term in~(\ref{e_su3ham})].
The quadrupole interaction
is in fact a combination of Casimir operators,
\begin{equation}
Q\cdot Q=
4C_2[{\rm SU}(3)]-3\bar L^2=
4C_2[{\rm SU}(3)]-3C_2[{\rm SO}(3)],
\label{su3cas}
\end{equation}
and it follows that the Hamiltonian~(\ref{e_su3ham})
has the eigenvalues
\begin{equation}
E(\lambda,\mu,L)=E_0-
g_2\left[
4(\lambda^2+\mu^2+\lambda\mu+3\lambda+3\mu)-3L(L+1)
\right],
\label{su3eig}
\end{equation}
where $E_0$ is a constant energy
associated with the first term in the Hamiltonian~(\ref{e_su3ham})
and $\lambda$ and $\mu$ label the SU(3) representations.
The quadrupole interaction represents an example of symmetry breaking
since the degeneracy associated with an entire oscillator shell
is lifted by the quadrupole interaction.

The importance of Elliott's idea
is that it gives rise
to a rotational classification of states
through mixing of spherical configurations.
With the SU(3) model it was shown, for the first time,
how deformed nuclear shapes
may arise out of the spherical shell model.
As a consequence, Elliott's work bridged the gap
between the nuclear shell model 
and the liquid droplet model
which up to that time (1958)
existed as separate views of the nucleus.

\section{The interacting boson model}
\label{s_ibm}
Arguably more than any other model of the nucleus,
the interacting boson model (IBM) illustrates
the power of group-theoretical techniques
and the physics insights that can be obtained from them.
In this section a brief introduction to the IBM is given
with the primary goal to provide an example
of the notion of dynamical symmetry
which was introduced in Sect.~\ref{s_dsym}.
It is not the aim here to give a full account of the IBM
which can be found in the book of Iachello and Arima~\cite{Iachello87}.

\subsection{The model}
\label{ss_ibm}
The building blocks of the IBM
are $s$ and $d$ bosons with angular momenta $\ell=0$ and $\ell=2$.
A nucleus is characterized by a constant total number of bosons $N$
which equals half the number of valence nucleons
(particles or holes, whichever is smaller).
In these lecture notes no distinction is made between neutron and proton bosons,
an approximation which is known as \mbox{IBM-1}.

Since the Hamiltonian of the \mbox{IBM-1}
conserves the total number of bosons,
it can be written in terms of the 36 operators $b_{\ell m_\ell}^\dag b_{\ell' m'_\ell}$
where $b_{\ell m_\ell}^\dag$ ($b_{\ell m_\ell}$) creates (annihilates)
a boson with angular momentum $\ell$ and $z$ projection $m_\ell$.
This set of 36 operators generates the Lie algebra U(6).
A Hamiltonian that conserves the total number of bosons
is of the generic form
\begin{equation}
H=E_0+H_1+H_2+H_3+\cdots,
\label{e_ibmham}
\end{equation}
where the index refers to the order of the interaction
in the generators of U(6).
The first term $E_0$ is a constant
which represents the binding energy of the core. 
The second term is the one-body part
\begin{equation}
H_1=
\epsilon_s[s^\dag\times\tilde s]^{(0)}+
\epsilon_d\sqrt{5}[d^\dag\times\tilde d]^{(0)}\equiv
\epsilon_sn_s+
\epsilon_dn_d,
\label{e_ibmham1}
\end{equation}
where $\times$ refers to coupling in angular momentum,
$\tilde b_{\ell m_\ell}\equiv(-)^{\ell-m_\ell}b_{\ell,-m_\ell}$
and the coefficients $\epsilon_s$ and $\epsilon_d$
are the energies of the $s$ and $d$ bosons.
The third term in the Hamiltonian~(\ref{e_ibmham})
represents the two-body interaction
\begin{equation}
H_2=
\sum_{\ell_1\leq\ell_2,\ell'_1\leq\ell'_2,L}
\tilde v^L_{\ell_1\ell_2\ell'_1\ell'_2}
[[b^\dag_{\ell_1}\times b^\dag_{\ell_2}]^{(L)}\times
[\tilde b_{\ell'_2}\times\tilde b_{\ell'_1}]^{(L)}]^{(0)}_0,
\label{e_ibmham2}
\end{equation}
where the coefficients $\tilde v$ are related to the interaction matrix elements
between normalized two-boson states, 
$$
\langle\ell_1\ell_2;LM|H_2|\ell'_1\ell'_2;LM\rangle=
\sqrt{\frac{(1+\delta_{\ell_1\ell_2})(1+\delta_{\ell'_1\ell'_2})}{2L+1}}
\tilde v^L_{\ell_1\ell_2\ell'_1\ell'_2}.
$$
Since the bosons are necessarily symmetrically coupled,
allowed two-boson states are
$s^2$ ($L=0$), $sd$ ($L=2$) and $d^2$ ($L=0,2,4$).
Since for $n$ states with a given angular momentum
one has $n(n+1)/2$ interactions,
seven independent two-body interactions $v$ are found:
three for $L=0$,
three for $L=2$
and one for $L=4$.

This analysis can be extended to higher-order interactions.
One may consider, for example, the three-body interactions
$\langle\ell_1\ell_2\ell_3;LM|H_3|\ell'_1\ell'_2\ell'_3;LM\rangle$.
The allowed three-boson states are
$s^3$ ($L=0$),
$s^2d$ ($L=2$),
$sd^2$ ($L=0,2,4$)
and $d^3$ ($L=0,2,3,4,6$),
leading to $6+6+1+3+1=17$ independent three-body interactions
for $L=0,2,3,4,6$, respectively.

\subsection{Dynamical symmetries}
\label{ss_ibmds}
The characteristics of the most general IBM Hamiltonian
which includes up to two-body interactions
and its group-theoretical properties
are by now well understood~\cite{Castanos79}.
Numerical procedures exist
to obtain its eigensolutions
but the problem can be solved analytically for particular choices
of boson energies and boson--boson interactions.
For an IBM Hamiltonian with up to two-body interactions
between the bosons,
three different analytical solutions or limits exist:
the vibrational U(5)~\cite{Arima76},
the rotational SU(3)~\cite{Arima78}
and the $\gamma$-unstable SO(6) limit~\cite{Arima79}.
They are associated with the algebraic reductions
\begin{equation}
{\rm U}(6)\supset
\left\{\begin{array}{c}
{\rm U}(5)\supset{\rm SO}(5)\\
{\rm SU}(3)\\
{\rm SO}(6)\supset{\rm SO}(5)
\end{array}\right\}
\supset{\rm SO}(3).
\label{e_ibmlat}
\end{equation}
The algebras appearing in the lattice~(\ref{e_ibmlat})
are subalgebras of U(6)
generated by operators of the type $b^\dag_{\ell m_\ell}b_{\ell'm'_\ell}$,
the explicit form of which is listed,
for example, in Ref.~\cite{Iachello87}.
With the subalgebras U(5), SU(3), SO(6), SO(5) and SO(3)
there are associated one linear [of U(5)]
and five quadratic Casimir operators.
The total of all one- and two-body interactions
can be represented by including in addition
the operators
$C_1[{\rm U}(6)]$,
$C_2[{\rm U}(6)]$
and $C_1[{\rm U}(6)]C_1[{\rm U}(5)]$.
The most general IBM Hamiltonian
with up to two-body interactions
can thus be written
in an {\em exactly} equivalent way with Casimir operators.
Specifically, the Hamiltonian reads
\begin{eqnarray}
H_{1+2}&=&
\kappa_1C_1[{\rm U}(5)]+
\kappa^\prime_1C_2[{\rm U}(5)]+
\kappa_2C_2[{\rm SU}(3)]
\nonumber\\&&+
\kappa_3C_2[{\rm SO}(6)]+
\kappa_4C_2[{\rm SO}(5)]+
\kappa_5C_2[{\rm SO}(3)],
\label{e_ibmhamc}
\end{eqnarray}
which is just an alternative way of writing
$H_1+H_2$ of Eqs.~(\ref{e_ibmham1},\ref{e_ibmham2})
if interactions are omitted
that contribute to the binding energy only.

The representation~(\ref{e_ibmhamc}) is much more telling
when it comes to the symmetry properties
of the IBM Hamiltonian.
If some of the coefficients $\kappa_i$ vanish
such that $H_{1+2}$
contains Casimir operators of subalgebras
belonging to a {\em single} reduction in the lattice~(\ref{e_ibmlat}),
then the eigenvalue problem can be solved analytically.
Three classes of spectrum generating Hamiltonians
can thus be constructed of the form
\begin{eqnarray}
{\rm U}(5)&:&H_{1+2}=
\kappa_1C_1[{\rm U}(5)]+
\kappa^\prime_1C_2[{\rm U}(5)]+
\kappa_4C_2[{\rm SO}(5)]+
\kappa_5C_2[{\rm SO}(3)],
\nonumber\\
{\rm SU}(3)&:&H_{1+2}=
\kappa_2C_2[{\rm SU}(3)]+
\kappa_5C_2[{\rm SO}(3)],
\nonumber\\
{\rm SO}(6)&:&H_{1+2}=
\kappa_3C_2[{\rm SO}(6)]+
\kappa_4C_2[{\rm SO}(5)]+
\kappa_5C_2[{\rm SO}(3)].
\label{e_ibmlim}
\end{eqnarray}
In each of these limits
the Hamiltonian is written as a sum of commuting operators
and, as a consequence, the quantum numbers
associated with the different Casimir operators are conserved.
They can be summarized as follows:
\begin{eqnarray}
&&\begin{array}{ccccccccc}
{\rm U}(6)&\supset&{\rm U}(5)&\supset&{\rm SO}(5)&
\supset&{\rm SO}(3)&\supset&{\rm SO}(2)\\
\downarrow&&\downarrow&&\downarrow&&\downarrow&&\downarrow\\[0mm]
[N]&&n_d&&\tau&&\nu_\Delta L&&M_L
\end{array},
\nonumber\\[1ex]
&&\begin{array}{ccccccc}
{\rm U}(6)&\supset&{\rm SU}(3)&\supset&{\rm SO}(3)&
\supset&{\rm SO}(2)\\
\downarrow&&\downarrow&&\downarrow&&\downarrow\\[0mm]
[N]&&(\lambda,\mu)&&K_LL&&M_L
\end{array},
\nonumber\\[1ex]
&&\begin{array}{ccccccccc}
{\rm U}(6)&\supset&{\rm SO}(6)&\supset&{\rm SO}(5)&
\supset&{\rm SO}(3)&\supset&{\rm SO}(2)\\
\downarrow&&\downarrow&&\downarrow&&\downarrow&&\downarrow\\[0mm]
[N]&&\sigma&&\tau&&\nu_\Delta L&&M_L
\end{array}.
\label{e_ibmclas}
\end{eqnarray}
Furthermore, for each of the three Hamiltonians in Eq.~(\ref{e_ibmlim})
an analytic eigenvalue expression is available,
\begin{eqnarray}
{\rm U}(5)&:&E(n_d,\tau,L)=
\kappa_1 n_d+
\kappa^\prime_1 n_d(n_d+4)+
\kappa_4 \tau(\tau+3)+
\kappa_5 L(L+1),
\nonumber\\
{\rm SU}(3)&:&E(\lambda,\mu,L)=
\kappa_2 (\lambda^2+\mu^2+\lambda\mu+3\lambda+3\mu)+
\kappa_5 L(L+1),
\nonumber\\
{\rm SO}(6)&:&E(\sigma,\tau,L)=
\kappa_3 \sigma(\sigma+4)+
\kappa_4 \tau(\tau+3)+
\kappa_5 L(L+1).
\label{e_ibmeig}
\end{eqnarray}
One can add Casimir operators of U(6)
to the Hamiltonians in Eq.~(\ref{e_ibmhamc})
without breaking any of the symmetries.
For a given nucleus they reduce to a constant contribution.
They can be omitted
if one is only interested in the spectrum of a single nucleus
but they should be introduced
if one calculates binding energies.
Note that none of the Hamiltonians in Eq.~(\ref{e_ibmlim})
contains a Casimir operator of SO(2).
This interaction breaks the SO(3) symmetry
(lifts the $M_L$ degeneracy)
and would only be appropriate if the nucleus is placed
in an external electric or magnetic field.

The dynamical symmetries of the IBM arise
if combinations of certain coefficients $\kappa_i$
in the Hamiltonian~(\ref{e_ibmhamc}) vanish.
The converse, however, cannot be said.
Even if all parameters $\kappa_i$ are non-zero,
the Hamiltonian $H_{1+2}$
still may exhibit a dynamical symmetry
and be analytically solvable.
This is a consequence of the existence
of unitary transformations
which preserve the eigenspectrum
of the Hamiltonian $H_{1+2}$
(and hence its analyticity properties)
and which can be represented as transformations
in the parameter space $\{\kappa_i\}$.
A {\em systematic} procedure exists
for finding such transformations
or parameter symmetries~\cite{Shirokov98}
which can, in fact, be applied
to any Hamiltonian
describing a system of interacting bosons and/or fermions.

While a numerical solution
of the shell-model eigenvalue problem
in general rapidly becomes impossible
with increasing particle number,
the corresponding problem in the IBM with $s$ and $d$ bosons
remains tractable at all times,
requiring the diagonalization of matrices
with dimension of the order of $\sim10^2$.
One of the main reasons for the success of the IBM
is that it provides a workable, albeit approximate, scheme
which allows a description of transitional nuclei
with a few parameters.

\subsection{Partial dynamical symmetries}
\label{ss_ibmpds}
As argued in Sect.~\ref{s_qm},
a dynamical symmetry can be viewed
as a generalization and refinement of the concept of symmetry.
Its basic paradigm is to write a Hamiltonian
in terms of Casimir operators of a set of nested algebras.
Its hallmarks are
(i) solvability of the complete spectrum,
(ii) existence of exact quantum numbers for all eigenstates
and (iii) pre-determined structure  of the eigenfunctions,
independent of the parameters in the Hamiltonian.
A further enlargement of these ideas is obtained
by means of the concept of partial dynamical symmetry.
The idea is to relax the conditions of {\em complete} solvability
and this can be done in essentially two different ways:
\begin{enumerate}
\item
{\it Some of the eigenstates keep all of the quantum numbers.}
In this case the properties of solvability, good quantum numbers,
and symmetry-dictated structure are fulfilled exactly,
but only by a subset of eigenstates~\cite{Alhassid92,Leviatan96}.
\item
{\it All eigenstates keep some of the quantum numbers.}
In this case none of the eigenstates is solvable,
yet some quantum numbers (of the conserved symmetries)
are retained.
In general, this type of partial dynamical symmetry arises
if the Hamiltonian preserves some of the quantum numbers
in a dynamical-symmetry classification
while breaking others~\cite{Leviatan86,Isacker99b}.
\end{enumerate}
Combinations of 1 and 2 are possible as well,
for example, if some of the eigenstates
keep some of the quantum numbers~\cite{Leviatan02}.

It should be emphasized that dynamical symmetry, be it partial or not,
is a notion that is not restricted to a specific model
but can be applied to any quantal system consisting of interacting particles.
Quantum Hamiltonians with a partial dynamical symmetry
can be constructed with general techniques 
and their existence is closely related
to the order of the interaction among the particles.
Applications of these concepts continue to be explored
in all fields of physics.

\subsection{Microscopy}
\label{ss_mic}
The connection of the IBM with the shell model
arises by identifying the $s$ and $d$ bosons
with correlated (or Cooper) pairs
formed by two nucleons in the valence shell
coupled to angular momentum $J=0$ and $J=2$.
There exists a rich and varied literature on general procedures
to carry out boson mappings
in which pairs of fermions are represented as bosons.
They fall into two distinct classes.
In the first one establishes a correspondence
between boson and fermion operators
by requiring them to have the same algebraic structure,
that is, the same commutation relations.
In the second class the correspondence is established
rather between state vectors in both spaces.
In each case further subclasses exist
that differ in their technicalities
({\it e.g.}, the nature of the operator expansion
or the hierarchy in the state correspondence).
In the specific example at hand,
namely the mapping between the IBM and the shell model,
arguably the most successful procedure
has been the so-called OAI mapping~\cite{Otsuka78b}
which associates vectors based
on a seniority [U(5)] hierarchy in fermion (boson) space.
It has been used in highly complex situations
that go well beyond the simple version of \mbox{IBM-1}
with just identical $s$ and $d$ bosons
and which include, for example,
neutron--proton $T=1$ and $T=0$ pairs~\cite{Thompson87,Juillet01}.

\subsection{The classical limit}
\label{ss_climit}
The connection of the IBM with the geometric model of the nucleus
can be obtained on the basis of coherent-state formalism~\cite{Ginocchio80,Dieperink80,Bohr80}.
The central outcome of the formalism is that for any \mbox{IBM-1} Hamiltonian
a corresponding potential $V(\beta,\gamma)$ can be constructed
where $\beta$ and $\gamma$ parametrize
the intrinsic quadrupole deformation of the nucleus~\cite{Bohr75}.
This procedure is known as the classical limit of the IBM.

The coherent states used for obtaining the classical limit of the IBM
are of the form
\begin{equation}
|N;\alpha_\mu\rangle\propto
\left(s^\dag+\sum_\mu\alpha_\mu d^\dag_\mu\right)^N|{\rm o}\rangle,
\label{e_coh}
\end{equation}
where $|{\rm o}\rangle$ is the boson vacuum
and $\alpha_\mu$ are five complex variables.
These have the interpretation of (quadrupole) shape variables
and their associated conjugate momenta.
If one limits oneself to static problems,
the $\alpha_\mu$ can be taken as real;
they specify a shape
and are analogous to the shape variables
of the droplet model of the nucleus~\cite{Bohr75}.
The $\alpha_\mu$ can be related to three Euler angles
which define the orientation of an intrinsic frame of reference,
and two intrinsic shape variables, $\beta$ and $\gamma$,
that parametrize quadrupole vibrations
of the nuclear surface around an equilibrium shape.
In terms of the latter variables,
the coherent state~(\ref{e_coh}) is rewritten as
\begin{equation}
|N;\beta\gamma\rangle\propto
\left(s^\dag+
\beta\left[\cos\gamma\,d^\dag_0
+\sqrt{\frac 1 2}\sin\gamma\,(d^\dag_{-2}+d^\dag_{+2})\right]
\right)^N|{\rm o}\rangle.
\label{e_cohb}
\end{equation}
The expectation value of the Hamiltonian~(\ref{e_ibmham}) in this state
can be determined by elementary methods~\cite{Isacker81}
and yields a functional expression in $\beta$ and $\gamma$
which is identified with a potential $V(\beta,\gamma)$,
familiar from the geometric model.
The classical limit of the most general Hamiltonian~(\ref{e_ibmham})
is found to be of the generic form
\begin{equation}
V(\beta,\gamma)=
E_0+\sum_{n\geq1}\frac{N(N-1)\cdots(N-n+1)}{(1+\beta^2)^n}
\sum_{kl}a^{(n)}_{kl}\beta^{2k+3l}\cos^l3\gamma,
\label{e_climit}
\end{equation}
where the coefficients $a^{(n)}_{kl}$
can be expressed in terms of the single-boson energies
and $n$-body interactions between the bosons.

A catastrophe analysis~\cite{Gilmore81}
of the potential surfaces in $(\beta,\gamma)$
as a function of the Hamiltonian parameters
determines the stability properties of these shapes.
This analysis was carried out for the general IBM Hamiltonian
with up to two-body interactions by L\'opez--Moreno and Casta\~nos~\cite{Lopez96}.
The results of this study are confirmed~\cite{Jolie01}
if a simplified IBM Hamiltonian is considered of the form
\begin{equation}
H_{1+2}=
\epsilon\,n_d +
\kappa\,Q\cdot Q.
\label{e_ibmhamcqf}
\end{equation}
This Hamiltonian provides a simple parametrization
of the essential features of nuclear structural evolution
in terms of a vibrational term $n_d$ (the number of $d$ bosons)
and a quadrupole interaction $Q\cdot Q$ with
\begin{equation}
Q_\mu=
[s^\dag\times\tilde d+d^\dag\times\tilde s]^{(2)}_\mu+
\chi[d^\dag\times\tilde d]^{(2)}_\mu.
\label{e_quad}
\end{equation}
Besides an overall energy scale,
the spectrum of the Hamiltonian~(\ref{e_ibmhamcqf})
is determined by two parameters: the ratio $\epsilon/\kappa$ and $\chi$.
The three limits of the IBM
are obtained with an appropriate choice of parameters:
U(5) if $\kappa=0$,
${\rm SU}_\pm(3)$ if $\epsilon=0$ and $\chi=\pm\sqrt{7}/2$,
and SO(6) if $\epsilon=0$ and $\chi=0$.
One may thus represent the parameter space
of the simplified IBM Hamiltonian~(\ref{e_ibmhamcqf}) on a triangle
with vertices that correspond to the three limits U(5), SU(3) and SO(6),
and where arbitrary points correspond to
specific values of $\epsilon/\kappa$ and $\chi$.
Since there are two possible choices for SU(3),
$\chi=-\sqrt{7}/2$ and $\chi=+\sqrt{7}/2$,
the triangle can be extended to cover both cases
by allowing $\chi$ to take negative as well as positive values. 

\begin{figure}
\centering
\includegraphics[width=7cm]{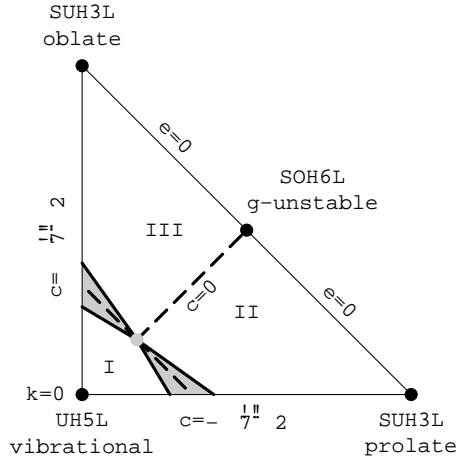}
\caption{
Phase diagram of the Hamiltonian~(\ref{e_ibmhamcqf})
and the associated geometric interpretation.
The parameter space is divided into three regions
depending on whether the corresponding potential
has (I) a spherical, (II) a prolate deformed or (III) an oblate deformed absolute minimum.
These regions are separated by dashed lines
and meet in a triple point (grey dot).
The shaded area corresponds to a region
of coexistence of a spherical and a deformed minimum.
Also indicated are the points on the triangle (black dots)
which correspond to the dynamical-symmetry limits
of the Hamiltonian~(\ref{e_ibmhamcqf})
and the choice of parameters $\epsilon$, $\kappa$ and $\chi$
for specific points or lines of the diagram.}
\label{f_trib}
\end{figure}

The geometric interpretation of any IBM Hamiltonian on the triangle
can now be found from its expectation value
in the coherent state~(\ref{e_cohb})
which for the particular Hamiltonian~(\ref{e_ibmhamcqf}) gives
\begin{eqnarray}
V(\beta,\gamma) &=&
\frac{N\epsilon\beta^2}{1+\beta^2} + 
\kappa\left[
\frac{N(5+(1+\chi^2) \beta^2)}{1+\beta^2} 
\right.
\nonumber\\
&&\left.
+\frac{N(N-1)}{(1+\beta^2)^2}
\left( \frac{2}{7}\chi^2\beta^4-
4\sqrt{\frac{2}{7}}\chi\beta^3\cos3\gamma+
4 \beta^2 \right) 
\right].
\label{potcqf}
\end{eqnarray}
The catastrophe analysis of this surface 
is summarized with the phase diagram shown in Fig.~\ref{f_trib}.
Analytically solvable limits are indicated by the dots.
Two different SU(3) limits occur
corresponding to two possible choices
of the quadrupole operator,
$\chi=\pm\sqrt{7}/2$.
Close to the U(5) vertex,
the IBM Hamiltonian has a vibrational-like spectrum.
Towards the SU(3) and SO(6) vertices,
it acquires rotational-like characteristics.
This is confirmed by a study
of the character of the potential surface in $\beta$ and $\gamma$
associated with each point of the triangle.
In the region around U(5),
corresponding to large $\epsilon/\kappa$ ratios,
the minimum of the potential is at $\beta=0$.
On the other hand,
close to the ${\rm SU}_+(3)$--SO(6)--${\rm SU}_-(3)$ axis
the IBM Hamiltonian corresponds to a potential
with a deformed minimum.
Furthermore, in the region around prolate ${\rm SU}_-(3)$ ($\chi<0$)
the minimum occurs for $\gamma=0^{\rm o}$
while around oblate ${\rm SU}_+(3)$ ($\chi>0$)
it does for $\gamma=60^{\rm o}$.
In this way the picture emerges that
the IBM parameter space can be divided into three regions
according to the character of the associated potential having
(I) a spherical minimum,
(II) a prolate deformed minimum
or (III) an oblate deformed minimum.
The boundaries between the different regions
(the so-called Maxwell set)
are indicated by the dashed lines in Fig.~\ref{f_trib}
and meet in a triple point.
The spherical--deformed border region
displays another interesting phenomenon.
Since the {\em absolute} minimum of the potential
must be either spherical, or prolate or oblate deformed,
its character uniquely determines the three regions
and the dividing Maxwell lines.
Nevertheless, this does not exclude the possibility
that, in passing from one region to another,
the potential may display a second {\em local} minimum.
This indeed happens for the U(5)--SU(3) transition~\cite{Iachello98}
where there is a narrow region of coexistence
of a spherical and a deformed minimum,
indicated by the shaded area in Fig.~\ref{f_trib}.
Since, at the borders of this region of coexistence,
the potential undergoes a {\em qualitative} change of character,
the boundaries are genuine critical lines
of the potential surface~\cite{Gilmore81}. 

Although these geometric results have been obtained
with reference to the simplified Hamiltonian~(\ref{e_ibmhamcqf})
and its associated `triangular' parameter space,
they remain valid for the general IBM Hamiltonian
with up to two-body interactions~\cite{Lopez96}.

\section{Summary}
\label{s_conc}
In these lecture notes an introduction was given
to the notions of symmetry 
and dynamical symmetry (or spectrum generating algebra).
Their use in the solution of the (nuclear) many-body problem
was described.
Two particular examples of these techniques
were discussed in detail:
(i) SO(4) symmetry of the hydrogen atom
and (ii) isospin symmetry in nuclei.
A review was given
of the shell model and the interacting boson model,
with particular emphasis on the application of group-theoretical techniques
in the context of these models.

\section*{Acknowledgment}
This paper is dedicated to the memory of Marcos Moshinsky,
the intellectual father of Mexican physics
and founder of the {\it Escuela Latino Americana de F\'\i sica}.
The two years I have spent in Mexico as a visitor
and the many hours with Marcos as a teacher,
were crucial to my formation as a physicist.
Without him I never could have given these lectures.

\end{document}